\documentclass[11pt,a4paper,pdftex]{article}
\pdfoutput=1
\usepackage{jcappub}

\newcommand{\be}{\begin{eqnarray}}
\newcommand{\ee}{\end{eqnarray}}

\begin{document}

\title{Can we observationally test the weak cosmic censorship conjecture?}

\author{Lingyao Kong, Daniele Malafarina and Cosimo Bambi$^1$ 
\note{Corresponding author}}

\emailAdd{071019019@fudan.edu.cn}
\emailAdd{daniele@fudan.edu.cn}
\emailAdd{bambi@fudan.edu.cn}

\affiliation{Center for Field Theory and Particle Physics \& Department of Physics,\\
Fudan University,\\ 
220 Handan Road, 200433 Shanghai, China}

\abstract{In general relativity, gravitational collapse of matter fields ends 
with the formation of a spacetime singularity, where the matter 
density becomes infinite and standard physics 
breaks down. According to the weak cosmic censorship conjecture, singularities 
produced in the gravitational collapse cannot be seen by distant observers and 
must be hidden within black holes. The validity of this conjecture is still controversial 
and at present we cannot exclude that naked singularities can be created in our 
Universe from regular initial data. 
In this paper, we study the radiation emitted by a collapsing cloud of 
dust and check whether it is possible to distinguish the birth of a black hole from 
the one of a naked singularity. In our simple dust model, we find that the properties 
of the radiation emitted in the two scenarios is qualitatively similar. That suggests 
that observational tests of the cosmic censorship conjecture may be very difficult, 
even in principle.}


\maketitle


\section{Introduction}

One of the most important open problems in gravitational physics is that 
of the final fate of a heavy star after exhausting its nuclear fuel. For normal stars,
the object contracts up to when the quantum pressure of electrons or
neutrons stops the collapse and the outcome is either a white dwarf or 
a neutron star. However, if the star is very massive, there is no known 
mechanism capable of compensating the inward push of its own gravitational force, and 
the body will undergo a complete gravitational collapse. According to the theory 
of general relativity, the final product of gravitational collapse must be a
spacetime singularity~\cite{hp,he}. In principle, the singularity may 
either be hidden behind a horizon, and in this case the result of the 
collapse is a black hole, or be naked, and thus visible to distant 
observers. While the weak cosmic censorship conjecture asserts that 
singularities created in gravitational collapse must be hidden within 
black holes~\cite{wccc}, today we know many physically relevant 
counterexamples in which naked singularities are formed from regular 
initial data (for a recent review, see e.g. Ref.~\cite{jm} and 
Ref.~\cite{Joshibook} for a detailed treatment). The possibility 
of detecting radiation from the high curvature region where classically 
we would expect the formation of a singularity would represent a unique 
opportunity to investigate strong gravity and observationally test the 
region where quantum gravity phenomena are supposed to show 
up~\cite{z2,bmm1,bmm2}.

The predictions of general relativity have been confirmed by experiments 
in Earth's gravitational field~\cite{llr,gpb}, by spacecraft missions in 
the Solar System~\cite{cassini}, and by accurate radio observations of 
binary pulsars~\cite{wt05,pulsar} (for a general review, see e.g. Ref.~\cite{will}). 
In all these environments, the gravitational field is weak, in the sense 
that one can write $g_{tt} = 1 + \phi$ with $|\phi| \ll 1$. In the last few years, 
there have been significant efforts and progresses to test the theory in 
the strong field regime, where the approximation $|\phi| \ll 1$ breaks down.
The ideal laboratory to test strong gravitational fields is the spacetime 
around astrophysical black hole candidates (see e.g. 
Refs.~\cite{image,jet1,jet2,qpo,cfm,iron,wh,cfmiron,torres,jp1,jp2}; 
for a review see~\cite{rev1,rev2}). 
These studies have shown that the properties of the electromagnetic radiation 
emitted by the gas in the accretion disk can provide useful details about the 
spacetime geometry around the compact object and available radio and X-ray 
data can already be used to constrain new physics.

While deviations from the predictions of general relativity in the spacetime
around astrophysical black hole candidates are definitively possible, since current 
observations can put only weak constraints, from purely theoretical arguments new 
physics is not strictly necessary here (see however~\cite{mathur,gd1,gd2,gd3}).
The black hole's event horizon has indeed no special properties for a
freely falling observer. On the contrary, the existence of spacetime singularities,
where observer-independent quantities like the scalar curvature or the 
Kretschmann scalar may diverge, is very likely a symptom of the break 
down of general relativity and new physics, presumably a quantum theory 
of gravity, is mandatory. Some observational tests have already been
proposed in the literature~\cite{ns1,ns2,ns3,ns4,ns5}.
In this paper we study the question of principle whether the high density 
region close to the formation of the singularity can affect the outside 
universe by exploring a toy model describing the radiation emitted from 
the high curvature region of astrophysical collapsing bodies, where 
classically we would expect the formation of a singularity.
More specifically, we want to figure out if -- at least in 
principle -- we can observationally distinguish the case in which the classical 
singularity that forms at the end of the collapse is not covered by the horizon 
from the case in which the horizon forms before the singularity.
If this were to be possible, we would in principle be able to experimentally 
test the weak cosmic censorship conjecture.

During the collapse, the density and the temperature of the object increase. 
Subnuclear reactions, otherwise strongly suppressed, become important 
and the collapsing star can emit a large amount of energy in several forms 
of radiation. The luminosity curve of this radiation clearly depends on the 
evolution of the gravitational collapse, setting the evolution of the increase 
in density and temperature at any layer of the body. For instance, the detection 
of neutrinos from supernovae may be used to probe the equation of state of 
matter at supernuclear densities~\cite{naka1,naka2}.  
Experiments to detect neutrinos from supernovae 
already exist and they are simply waiting for the explosion of a nearby 
supernova. In 1987, neutrinos from the supernova SN1987A in the Large 
Magellanic Cloud were detected by experiments searching the proton decay 
(Kamiokane II, IMB, Baksan detected together 24 events). With the sensitivity 
of present experiments, a supernova explosion in our Galaxy could produce
thousands of events in a detector like Super-Kamiokande and even millions 
of events in one like IceCube. In the same way, some 
weakly interacting radiation may be used to track the gravitational collapse of 
an astrophysical body and observationally test if the collapse follows the 
pattern expected for the formation of a black hole, for the creation of a naked
singularity, or another one.

The simplest exact solution for gravitational collapse in which the outcome can be 
either a black hole or a naked singularity is the Leimatre-Tolman-Bondi (LTB)
dust model~\cite{lem,tol,bon}. The system has spherical symmetry and, 
depending on the initial density and velocity profile, it may behave in two different ways.
Either the horizon develops first and the subsequent singularity is always covered 
or, vice versa, a singularity visible to distant observers forms before the formation of
the horizon~\cite{dust1,dust2,dust3,dust4,dust5}. 
If we consider a distant observer and we integrate backwards in time the photons' trajectories, from the observer to the collapsing object, we can obtain the luminosity image at 
any time. Integrating over the whole image, we can find the curve luminosity 
produced by the collapsing object. As the evolution of the emissivity of the 
matter in the star depends on the gravitational collapse, the curve luminosity 
provide information on the collapse itself. In particular, one may expect that
the formation of a event horizon and of a naked singularity can be characterized
by qualitatively different light curves.

Unfortunately, this does not seem to be the case. Assuming two different emissivity 
functions, we find that both the black hole and the naked singularity case show
very similar luminosity spectrum. 
This would leave the distinction undetermined even
once a more realistic scenario is considered.
While the results shown here are based on a very simple analytical toy model, 
they definitely suggest that any observational test of the weak cosmic censorship 
conjecture through the 
analysis of the emitted spectrum of a collapsing astrophysical body may be 
extremely challenging unless some effects coming from new physics at high densities 
do not intervene to completely change the picture.

The content of our manuscript is organized as follows. In Section~\ref{s-1},
we briefly review the LTB dust collapse model. In Section~\ref{s-2}, we
describe how our code computes the luminosity of the collapsing object 
seen by a distant observer. In Section~\ref{s-3}, we present the results of
our numerical calculations. At first we consider the homogeneous case, 
the well-known Oppenheimer and Snyder model~\cite{OS}, in which the 
final product of the collapse is always a black hole. Then we extend the 
study to the inhomogeneous case, in which the collapse can create either 
a black hole or a naked singularity, depending on the initial matter density 
profile. We then compare the curve luminosity of the two scenarios. 
Summary and conclusions are reported in Section~\ref{s-c}. 
Throughout the paper, we use units in which
$G_{\rm N} = c = 1$.

\section{LTB dust collapse model \label{s-1}}

The LTB model describes a spherically symmetric system composed of  non 
interacting particles (dust) that undergoes complete gravitational collapse. 
The most general spherically symmetric line element can be written in
comoving coordinates (namely coordinates attached to the infalling particles) as
\begin{eqnarray}
ds^2 = -e^{2\nu}dt^2 + \frac{\rho'^2}{G}dr^2 
+ \rho^2 \left(d\theta^2 + \sin^2\theta d\phi^2\right) \, ,
\end{eqnarray}
where $\nu$, $\rho$, and $G$ are 
functions of the comoving time $t$ and radius $r$. 
Here and in what follows, the prime $(')$ denotes a derivative with respect 
to $r$. If we impose that $\nu$, $\rho$, and $G$ are independent of the $t$ 
coordinate, we find the class of static interior Schwarzschild solutions that was originally 
studied by Tolman
\cite{Tolman}.
As we are using comoving coordinates, the energy momentum tensor of a relativistic fluid 
is diagonal and can be written as
\begin{eqnarray}
T^\mu_\nu = {\rm diag}\{\epsilon(r,t), p_r(r,t), p_\theta(r,t), p_\theta(r,t) \} \, ,
\end{eqnarray}
where $\epsilon$ is the energy density and $p_r$ and $p_\theta$ are, 
respectively, the radial and tangential pressure of the cloud. 
Einstein's equations then take the form
\begin{eqnarray}
\epsilon &=& \frac{F'}{\rho^2 \rho'} \, , \label{ee-e} \\
p_r &=& - \frac{\dot{F}}{\rho^2 \dot{\rho}} \, , \label{ee-pr} \\
\nu' &=& 2 \frac{p_\theta - p_r}{\epsilon + p_r} \frac{\rho'}{\rho}
- \frac{p_r'}{\epsilon+p_r} \, , \label{ee-nu} \\
\dot{G} &=& \frac{2 \nu'}{\rho'} \dot{\rho} G  \label{ee-g} \, ,
\end{eqnarray}
where the dot $(\dot{})$ denotes a derivative with respect to $t$ and
$F$ is the Misner-Sharp mass, defined by
\begin{eqnarray}\label{misner}
F = \rho (1 - G + e^{-2\nu} \dot{\rho}^2) \, .
\end{eqnarray}
$F$ is proportional to the ``gravitational mass'' enclosed within the shell of radial
coordinate $r$ at the time $t$. 
Eventually, we have a set of five coupled first order differential equations in seven 
unknown functions of $r$ and $t$. In general, it may be hard or impossible to solve the system.
The usual approach is to look for simplifications that,
while preserving the physical features of interest, allow us to solve the equations and
say something about the nature of the collapse.

The dust case is obtained for particles that carry no self-interacting energy and can be described imposing that $p_r = p_\theta = 0$. It turns out that this case is a particularly simple model where
the set of Einstein's equations can be solved explicitly. From Eq.~\eqref{ee-nu}, we get
$\nu = \nu(t)$ and we can change the time gauge in order to have a $t$ coordinate for which $\nu = 0$. 
Then Eq.~\eqref{ee-pr} reduces to $\dot{F} = 0$, which implies $F=F(r)$; that is, 
the amount of matter enclosed within the shell labeled by $r$ is conserved during 
the collapse. This, in turn, means that there is no inflow or outflow of matter/energy during the collapse and therefore, given the spherical symmetry of the system, the matching with the exterior geometry can be done with the simple and well-known Schwarzschild spacetime
\cite{matching}. 
Furthermore, given the absence of pressures, the boundary radius $r_b$, which corresponds to the shrinking boundary area-radius $\rho_b(t)=\rho(r_b,t)$, can be chosen at will.
From the matching conditions, it is easy to see that 
$F(r_b) = 2M_{\rm Sch}$, where $M_{\rm Sch}$ is the mass parameter 
in the exterior Schwarzschild metric. 
From Eq.~\eqref{ee-g}, that for dust reads $\dot{G}=0$, we can obtain
$G=G(r)$ as a free function, which is convenient to write in the form $G(r)=1+f(r)$. 
Finally, Eq.~\eqref{misner} becomes the equation of motion of the system
\begin{eqnarray}\label{e-rhodot}
\dot{\rho}=-\sqrt{\frac{F}{\rho} + f} \, ,
\end{eqnarray}
with the minus sign necessary to describe collapse. Given a certain mass profile $F$, after choosing the free function $f$, we can integrate Eq.~\eqref{e-rhodot} to get $\rho(t,r)$. 
Plugging this solution into Eq.~\eqref{ee-e}, we obtain $\epsilon(r,t)$, thus
completely solving the system.

The free function $f$ coming from the integration of Eq.~\eqref{ee-g} is related to the velocity of the infalling particles. The collapse is said to be bound if $f < 0$, marginally bound if $f = 0$, and unbound if $f > 0$. 
In the rest of the manuscript, we will restrict our attention on the marginally
bound case $f = 0$, which represents particles that would have zero initial velocity at spatial infinity. 
The line element for the collapsing interior reduces to
\begin{eqnarray}
ds^2_{\rm int} = - dt^2 + \rho'^2dr^2 
+ \rho^2 \left(d\theta^2 + \sin^2\theta d\phi^2\right) \, .
\end{eqnarray}
The collapse process leads eventually to the formation of a 
black hole when all the matter passes the threshold of trapping 
surfaces located at the event horizon in the Schwarzschild 
exterior. The condition for the formation of trapped surfaces 
for the collapsing cloud is given by $1 - F/\rho = 0$, and 
it reduces to $1 - 2 M_{\rm Sch}/R = 0$, where $R$ is 
the Schwarzschild radial coordinate, in the static case in vacuum.  
All the matter falls into the spacetime singularity that forms at $r=0$ and it is
easy to check, for instance by evaluating the Kretschmann scalar, that this is a
true curvature singularity.
A factor of crucial importance for black hole physics is to determine whether 
the singularity always forms at a later time with respect to the formation of the horizon,
thus being hidden to far away observers at all times, or if other possibilities 
are allowed.

The whole system has a gauge degree of freedom
that can be fixed by setting the scale at a certain time. In models of collapse,
one usually sets the area radius $\rho(t,r)$ equal to the comoving radius $r$
at the initial time $t_{i}$; that is, $\rho(t_{i},r)=r$. We can then introduce a scale
function $a$ as
\begin{eqnarray}
\rho(t,r)=ra(t,r) \, .
\end{eqnarray} 
with the initial condition $a(t_{i},r)=1$.
Further one wishes to impose certain regularity conditions to ensure the 
physical validity of the model. For example, one wishes to have a density 
profile that is regular at the center at the initial time and that presents no 
cusps in $r=0$ at all times. In order to have such regularity conditions, 
we can impose that the mass function near the center behaves in a suitable 
way. Therefore we can define a function $M(r)$ such that
\begin{equation}
    F(r)=r^3M(r) \, .
\end{equation}
We can rewrite the whole system of equations in terms of $M$ and $a$ and 
it is immediately found that the form of Eq.~\eqref{e-rhodot} with $f=0$ is invariant 
under the substitution of $F$ with $M$ and $\rho$ with $a$.
The energy density can now be written as
\begin{equation}\label{dens}
    \epsilon(r,t)=\frac{3M+rM'}{a^2(a+ra')} \, ,
\end{equation}
and it is easy to check that the singularity occurs for $a=0$, while values of the 
central shell $r=0$ with $a\neq 0$ are regular. This solves the problem of the 
divergence of $\epsilon$ at $\rho=0$ at all times coming from Eq.~(\ref{ee-e}). 

Note from Eq.~\eqref{dens} that the density diverges also when $\rho'=a+ra'$ goes to zero.
This is related to the occurrence of so called ``shell crossing'' singularities
~\cite{cross1,cross2,cross3}. These are singularities that arise from a breakdown of the coordinate system rather than true gravitational singularities and they can generally be avoided by a suitable change of coordinates. Nevertheless it is important to check under what circumstances they can arise in the model in order to impose some prescription to rule them out.
It is not difficult to see that for dust collapse, if one imposes a decreasing energy profile, then no shell crossing occurs.
With the new scaling, together with the requirement of avoidance of shell crossing, the density diverges only when the singularity is achieved.

\subsection{Oppenheimer-Snyder collapse}

If, for simplicity, we want to describe homogeneous collapse where 
$\epsilon=\epsilon(t)$, we need to take $M(r)=M_0$, which implies $a=a(t)$.
Then $\epsilon=3M_0/a^3$ and the equation of motion reduces to
\begin{equation}\label{motion}
    \dot{a}=-\sqrt{\frac{M_0}{a}} \, .
\end{equation}
Eq.~(\ref{motion}) can be easily integrated. The solution is given by
\begin{equation}
a(t)=\left(1-\frac{3}{2}\sqrt{M_0}t\right)^{2/3} \, ,
\end{equation}
with initial time $t_{i}=0$. The singularity forms at the time $t_s=2/3\sqrt{M_0}$.
Here all the shells become singular at the same time and the horizon forms at the 
boundary before the formation of the singularity, therefore leaving the singularity 
always covered.

The boundary of the cloud collapses along the curve $\rho_b(t)=\rho(r_b,t)=r_b a(t)$ 
and the whole cloud becomes trapped inside the event horizon at the time 
$t_{tr}<t_s$ for which $\rho_b(t_{tr})=2M_{\rm Sch}=r_b^3M_0$, so 
\begin{equation}
t_{tr}=t_s-\frac{4M_{\rm Sch}}{3} \, .
\end{equation}
For the homogeneous dust collapse we thus have the formation of a spacelike simultaneous
singularity that is always covered by the horizon.
This means that the region of extremely high density at the center of the collapsing cloud close to the formation of the singularity is causally disconnected from the outside universe.

\subsection{Inhomogeneous collapse}

If we wish to analyze a more general case, still within the dust scenario, we can take $M(r)$ to be written as an expansion close to the center as
\begin{eqnarray}
M(r)&=&M_0+M_1r+M_2r^2+... 
\end{eqnarray}
In this case, the density profile $\epsilon(r,t)$ is not homogeneous any more and the mass profile $M(r)$ will determine its form.
It is easy to see that the behavior of trapped surfaces and the structure of formation of the singularity can change drastically
(see for example~\cite{ns1b,ns2b,ns2b-2,ns3b} and references therein).
It turns out that it is the sign of $M_1$ that will determine the character of the singularity curve and the apparent horizon near the center. If we require $M_1=0$ (as it is often done in astrophysical scenarios, where one desires to have only quadratic terms in the density and pressures), the behavior of the apparent horizon and of the singularity curve near the center will be determined by the value of $M_2$. In the following we will therefore consider $M_1=0$ and $M_2<0$. The case $M_2>0$ is not physically relevant, as it implies a density increasing with the radius.

Mathematically, since in the dust collapse there are no pressures, the matching with the outside region can be done at any radius and therefore, if one shows that the singularity is locally naked (meaning that there are outgoing geodesics originating at the singularity and reaching a finite radius without being trapped), then one can choose $r_b$ for the matching in order to make it globally naked (meaning that such geodesics can reach observers at infinity). In a realistic scenario, things might be different (see for example \cite{MJH}) and when pressures are considered it is preferable to perform the matching with the exterior region at the radius where the pressure vanishes. 
Nevertheless, the possibility remains that the central singularity be visible to far away observers
(see for example \cite{JMS}).

The above formalism is enough to obtain the necessary information about the behavior of the dust cloud close to $r=0$ and close to the formation of the singularity. Two scenarios are possible:
\begin{enumerate}
\item In the black hole case, the trapped surfaces form at an outer shell before the formation of the singularity. Essentially each shell $r>0$ becomes trapped at a time smaller than the time of formation of the central singularity, $t_s(0)$.
\item In the naked singularity case, the trapped surfaces form at the center at the time of formation of the singularity. This means that the shell $r=0$ becomes trapped at the time $t_s(0)$, and light rays escaping from the central singularity can reach far away observers. At later times the trapped surface expands, thus covering the singularity.
\end{enumerate}

The equation of motion is given by Eq.~\eqref{motion} with $M(r)$ in place of $M_0$ and the solution takes again the form
\begin{equation}\label{a}
a(r,t)=\left(1-\frac{3}{2}\sqrt{M(r)}t\right)^{2/3} \: .
\end{equation}
The singularity curve is described by the condition that $a(t_s(r),r)=0$ which gives
\begin{eqnarray}\label{ts2}
t_{s}(r)=\frac{2}{3\sqrt{M(r)}} \, ,
\end{eqnarray}
from which we can see that different shells become singular at different times, with the central shell becoming singular first in the case that $M_2<0$.
The apparent horizon curve is given by
\begin{eqnarray}\label{ts2}
t_{ah}(r)= t_{s}(r) - \frac{2}{3}r^3M(r)\, ,
\end{eqnarray}
and it is easy to check that $t_{ah}$ is also increasing 
from the center and that $t_{ah}(0)=t_{s}(0)$.
Therefore, in the inhomogeneous dust case with $M_2<0$ the central singularity 
is not trapped at the time of its formation and it may become trapped only afterwards. 
This means that for certain choices of the boundary radius 
the high density region that develops close to the singularity is causally connected to the outside universe
(see for example \cite{ns1b,ns2b,ns2b-2} and \cite{ortiz} for the complete conformal structure
of the model). Such a collapsing cloud can potentially bear an observational signature different from that of the black hole case discussed above. Not every negative value of $M_2$ is allowed. From the condition that the energy density is positive throughout the cloud we get the constraint
\begin{equation}
M_2 > - \frac{3}{5} \frac{M_0}{r_b^2} \, .
\end{equation}

The gravitational collapse of a dust cloud is just a simple toy model that has the advantage that it can be treated analytically. Obviously, if one wishes to describe a star, pressures are important\footnote{Though it has been suggested that matter might approach a dust-like behavior close to the formation of the singularity where very strong gravitational fields are present (essentially particles falling in close to the speed of light are not able to interact) \cite{Penrose2}.}. 
Of course here we are investigating a mathematical toy model describing a simple light spectrum emitted from the vicinity of a naked singularity that has no resemblance to the real spectrum emitted by a realistic collapsing object. However this investigation is important in that it helps us answer the question of principle of whether the visibility of the region surrounding the singularity could have long range effects in such a way as to make it distinguishable from the formation of a black hole.
Therefore these models can constitute a first step to investigate what could possibly be observationally detectable if such naked singularities happened in realistic star collapse or in the formation of supermassive compact objects.
In fact, if one thinks about supermassive compact objects, the formalism is exactly the same, but the time scales are much longer.
Actually, in this case the dust model could be a better approximation than in the star collapse case, since supermassive objects are less dense than stars and tracing of geodesics inside a supermassive collapsing object in more meaningful.

\subsection{Exterior geometry}

The whole spacetime can be described by a collapsing interior given by the LTB metric (in comoving coordinates $\{t,r\}$) that matches at the boundary $r_b$ to a vacuum exterior given by the Schwarzschild metric (in Schwarzschild coordinates $\{T,R\}$)
\begin{equation}
ds^2_{\rm out}=-\left(1-\frac{2M_{\rm Sch}}{R}\right)dT^2+\left(1-\frac{2M_{\rm Sch}}{R}\right)^{-1}dR^2+R^2d\Omega^2 \; .
\end{equation}
The two manifolds match across the 3-surface $\Sigma$ given by $r=r_b$, which corresponds to $\rho=\rho_b(t)=\rho(r_b,t)$, in the interior and $R=R_b(T)$ in the exterior.
The matching of the two manifolds is achieved imposing the continuity of the first and second fundamental forms across the surface.
Continuity of the first fundamental form reduces to continuity of the metric across the surface. The induced metric on the surface can be written as
\begin{eqnarray}
ds^2_\Sigma&=&-dt^2+\rho_b(t)^2d\Omega^2=\nonumber\\
&=& - \left(1-\frac{2M_{\rm Sch}}{R_b}\right)dT^2+\left(1-\frac{2M_{\rm Sch}}{R_b}\right)^{-1}
\left(\frac{dR_b}{dT}\right)^2 dT^2+R_b^2d\Omega^2 \, ,
\end{eqnarray}
and the coordinate transformation on the boundary of the cloud $\Sigma$ 
that relates $t$ to $T$ is given by 
\begin{equation}\label{time}
\frac{dt}{dT}=\sqrt{\left(1-\frac{2M_{{\rm Sch}}}{R_b}\right)-\left(1-\frac{2M_{{\rm Sch}}}{R_b}\right)^{-1}\left(\frac{dR_b}{dT}\right)^2} \, .
\end{equation}
The metric components are continuos across $\Sigma$ and we can 
express the trajectory of the boundary as $\rho_b(t)=R_b(T(t))$.

\section{Tracing photons \label{s-2}}

In this section we discuss the theoretical aspects of the procedure by which 
we intend to trace light rays from the collapsing cloud to far away observers.
We consider the geodesics starting at some far away initial radius $R_*$ 
at the time $T=\bar{T}$ so that $R(\bar{T})=R_*$. Then we follow $R(T)$ 
tracing the photon backwards in time along the path from the observer to 
the collapsing cloud. Three scenarios are possible:
\begin{enumerate}
\item The photon escapes to infinity never hitting either the cloud or the event horizon.
\item The photon hits the event horizon. 
\item The photon hits the collapsing cloud, thus reaching the boundary $R_b(T)$ at a time antecedent the formation of the horizon. In this case, the photon can either escape from the cloud, thus crossing again the boundary at a later time, or hit the event horizon.
\end{enumerate}

First of all, we consider the Schwarzschild solution to construct the image of the object for a far away observer. This is the image seen after that all the photons coming from the LTB region reached the observer.
Then we will consider the LTB region describing the collapsing cloud. This evolves from the initial time $T_i=T(t_i)$ until the formation of the event horizon at $T_{tr}=T(t_{tr})$ after which, from the perspective of external observers, we are left with a black hole. Therefore the image seen by the far away observer will change in time from the initial moment until the formation of the horizon.

Given the spherical symmetry of the spacetime, motion happens always on a plane and we can restrict our analysis to the equatorial plane without any loss of generality. Null geodesics are then described by the equation
\begin{equation}
\left(\frac{dR}{d\lambda}\right)^2=E^2-\frac{L^2}{R^2}\left(1-\frac{2M}{R}\right) \; ,
\end{equation}
for the Schwarzschild exterior and by
\begin{eqnarray}\label{ng}
&&\left(\frac{dt}{d\lambda}\right)^2
-\rho'\left(\frac{dr}{d\lambda}\right)^2-\frac{L^2}{\rho^2}=0 \; , \nonumber\\
&&\frac{d^2r}{d\lambda^2} 
+ 2\frac{\dot{\rho}'}{\rho'}\frac{dr}{d\lambda}
\frac{dt}{d\lambda}
+ \frac{\rho''}{\rho'}\left(\frac{dr}{d\lambda}\right)^2
- \frac{L^2}{\rho^3\rho'} = 0 \; ,
\end{eqnarray}
for the LTB interior. Here $\lambda$ is an affine parameter, while $E$ and $L$ are the conserved energy and angular momentum related to the killing vectors as defined below.
Once the photon hits the boundary of the cloud, as in the case (3) above, to trace it inside the cloud we have to change coordinates from $T$ to $t$ via Eq.~\eqref{time} and follow the geodesic $r(t)$ (using the comoving time $t$ as the affine parameter) obtained by solving equation \eqref{ng} with the same value for $L$ as the one used for the outer part of the trajectory
\footnote{If we restrict to the case of radial null geodesics ($L=0$) we get $\frac{dt}{dr}=\pm\rho'$
where the plus sign denotes outgoing geodesics, while the minus sign is for ingoing geodesics.
Then the problem of studying the behavior of radial null geodesics emanating from the center of the dust cloud translates into the Cauchy problem given by Eq.~\eqref{ng} with the initial value $t(0)=t_0$, where $t_0\in[t_i,t_s(0)]$.}.

\subsection{Geodesics in the Schwarzschild spacetime}

The Schwarzschild spacetime is static and spherically symmetric. We can thus 
define two quantities conserved along geodesics. They are related to 
the killing vectors associated to time translations and rotations. These quantities
are the energy $E$ and angular moment $L$ and are given by
\begin{eqnarray}
E=\left(1-\frac{2M_{{\rm Sch}}}{R}\right)\frac{dT}{d\lambda} \; , 
\quad L=R^{2}\frac{d\phi}{d\lambda} \; .
\end{eqnarray}
Since the trajectory of a photon is independent of its energy $E$, for the
study of null geodesics it is more convenient to use the ``impact parameter''
$b = L/E$ instead of $E$ and $L$. All the equations depend now on $b$ and
$E$ and $L$ never appear. From the expression of the Schwarzschild metric, 
we can write the equation for null geodesics as
\begin{equation}
\left(\frac{dR}{dT}\right)^{2}=\left(1-\frac{2M_{{\rm Sch}}}{R}\right)^{2}-\left(1-\frac{2M_{{\rm Sch}}}{R}\right)^{3}\frac{b^{2}}{R^{2}} \; .
\end{equation}
which, once integrated with the initial condition $R(\bar{T})=R_*$,
gives the trajectory $R(T)$ of the photon in the exterior spacetime.

The other ingredient necessary to trace the photon backward in time 
from the observer to the cloud 
is the trajectory of the boundary of the cloud as given by $R_{b}\left(T\right)$.
This allows us to determine whether and when the photon hits the boundary
of the collapsing object. From Eq.~\eqref{time} we use the equation of 
motion~\eqref{motion} written at the boundary as
\begin{equation}
\frac{d\rho_{b}}{dt}=-\sqrt{\frac{F}{\rho_{b}}}=-\sqrt{\frac{2M_{{\rm Sch}}}{R_{b}}}\; ,
\end{equation}
and, noting that
at the boundary of the collapsing object is $\rho_{b}(t)=R_{b}(T(t))$, we get
\begin{equation}
\frac{dR_{b}}{dT}=\frac{dR_{b}}{dt}\frac{dt}{dT}=\frac{d\rho_{b}}{dt}\frac{dt}{dT} \; .
\end{equation}
Now making use of equation~\eqref{time} we obtain
\begin{equation}
\frac{dR_{b}}{dT}=-\sqrt{\frac{2M_{{\rm Sch}}}{R_{b}}}\left(1-\frac{2M_{{\rm Sch}}}{R_{b}}\right)\label{eq:RbT}\; ,
\end{equation}
that, once integrated, gives
\begin{eqnarray}
T (R_{b}) & = & T_{0}
-\frac{2}{3} \frac{R_{b}^{3/2}}{\sqrt{2M_{{\rm Sch}}}}
-2\sqrt{2M_{{\rm Sch}} R_{b}}  + 2M_{{\rm Sch}} \ln
\left(\sqrt{R_b} +\sqrt{2M_{{\rm Sch}}}\right) 
+ \nonumber\\ && 
-2M_{{\rm Sch}}\ln\left(\sqrt{R_b}-\sqrt{2M_{{\rm Sch}}}\right)\; ,
\end{eqnarray}
that can be inverted to obtain $R_{b}(T)$. The intersection of the null geodesic $R(T)$ with the boundary curve $R_{b}(T)$ then gives the time $\bar{\bar{T}}$ at which the photon traveling along the geodesics hits the boundary.

\subsection{Geodesics in the LTB spacetime}

Some of the photons that are traced back in time from the screen will hit the boundary of the cloud and propagate in the interior. These are the actual photons that are coming from the collapsing object and we need to evaluate geodesics in the LTB interior in order to trace them.
These were first studied in
\cite{nakao}. 
The null geodesics in the interior can be obtained from
\begin{eqnarray}
\frac{dt}{d\lambda} &=& \sqrt{\rho'^2 
\left(\frac{dr}{d\lambda}\right)^2
+ \frac{b^2}{\rho^2}} \; , \label{eq:geo_ltb_t} \\
\frac{d^2r}{d\lambda^2} &=& 
-2\frac{\dot{\rho}'}{\rho'}
\sqrt{\rho'^2 \left(\frac{dr}{d\lambda}\right)^2 + \frac{b^2}{\rho^2}}\frac{dr}{d\lambda}
- \frac{\rho''}{\rho'}\left(\frac{dr}{d\lambda}\right)^2
+ \frac{b^2}{\rho^3\rho'} \; .
\label{eq:geo_ltb_r}
\end{eqnarray}
The photon now propagates inside the cloud following the null geodesics with initial conditions at the boundary, namely the photon motion in the LTB interior starts with position $\rho_b=R_b(\bar{\bar{T}})$ at the time $\bar{\bar{t}}=t(\bar{\bar{T}})$.
From the expression for $\rho=ra$ with $a$ given by equation~\eqref{a}, we can find the time at which the photon coming from the observer's screen is at the boundary in the interior coordinates. We have
\begin{eqnarray}
\frac{dR_b}{d\lambda} & = & -\sqrt{\frac{2M_{{\rm Sch}}}{R_b}}\frac{dt}{d\lambda}
+R_b'\frac{dr}{d\lambda}\label{eq:dR} \; , \\
\left(\frac{dT}{d\lambda}\right)^{2} & = & 
\left(1-\frac{2M_{{\rm Sch}}}{R_b}\right)^{-2}
\left[\left(\frac{dR_b}{d\lambda}\right)^{2}
+\left(1-\frac{2M_{{\rm Sch}}}{R_b}\right)\frac{b^{2}}{R_b^{2}}\right] \; .
\end{eqnarray}
Now, with the help of Eq.~(\ref{ng}), we get
\begin{eqnarray}
\frac{dT}{d\lambda} = 
\left( 1-\frac{2M_{{\rm Sch}}}{R} \right)^{-1}
\left(\frac{dt}{d\lambda}-R'\sqrt{\frac{2M_{{\rm Sch}}}{R}}
\frac{dr}{d\lambda}\right)\label{eq:dT} \; ,
\end{eqnarray}
and eventually we arrive at the initial conditions at the boundary for the first
derivatives of $t(\lambda)$ and $r(\lambda)$, namely
\begin{eqnarray}
\frac{dt}{d\lambda} & = & \frac{dT}{d\lambda}+\frac{\sqrt{\frac{2M_{{\rm Sch}}}{R}}\frac{dR}{d\lambda}}{\left(1-\frac{2M_{{\rm Sch}}}{R}\right)} \; ,
\label{eq:dt_init} \\
\frac{dr}{d\lambda} & = & \frac{1}{\rho'}\sqrt{\left(\frac{dt}{d\lambda}\right)^{2}-\frac{b^{2}}{R^{2}}}
\; . \label{eq:dr_init}
\end{eqnarray}
With the above equations, it is straightforward to numerically calculate all the photon trajectories.

\subsection{Observed spectrum}

In the following, we consider two examples, namely the homogenous collapse model, where we set $M=M_0$, and the inhomogeneous one, where we have $M=M_0+M_2r^2$.
The spectrum at the time $T$ measured by the distant observer is given
by~\cite{worm}
\begin{equation}
I\left(T,\nu_{{\rm obs}}\right)=\int2\pi b \, db\int_{\gamma}g^{3}jdl \; ,
\end{equation}
where $\gamma$ is the photon's path, $j$ is the emissivity per unit
volume in the rest frame of the emitter and $g$ is the gravitational
redshift
\begin{equation}
g=\frac{\nu_{{\rm obs}}}{\nu_{e}}=
\frac{k_{\mu}v_{{\rm obs}}^{\mu}}{k_{\nu}v_{e}^{\nu}}
=\frac{E}{\frac{dt}{d\lambda}} \; ,
\end{equation}
$\nu_{{\rm obs}}$ is the photon frequency as measured by the distant
observer, $\nu_{e}$ is the photon frequency with respect to the emitter,
$v_{{\rm obs}}^{\mu}=\left(1,0,0,0\right)$ is the 4-velocity of the
distant observer, $v_{e}^{\mu}=\left(1,0,0,0\right)$ is the 4-velocity
of the emitter, and $k^{\mu}$ is the 4-momentum of the photon. Also $dl$
is the proper length in the rest-frame of the emitter and in our model
it turns out to be equal to $dt$ 
\begin{equation}
dl=\sqrt{^{3}g_{ij}\frac{dx^{i}}{d\lambda}\frac{dx^{j}}{d\lambda}}d\lambda=dt\,.
\end{equation}
In the next section, for the sake of clarity, we will use two simple emissivity functions.
In the first model, we assume that the emission is monochromatic with rest-frame 
frequency $\nu_{\star}$ and proportional to the square of the energy 
density (as we may expect in a two-body collision)
\begin{equation}
\label{eq-delta}
j=\epsilon^{2} \delta\left(\nu_{e}-\nu_{\star}\right) \; .
\end{equation}
In the second example, we replace the monochromatic emission with an 
exponential function that could somehow mimic a thermal emission
(even if, strictly speaking, our object is made of dust and therefore the
temperature is zero)
\begin{equation}
\label{eq-exp}
j=\epsilon^{2}\nu_{e}^{2}\exp\left(-\frac{\nu_{e}}{\epsilon^{1/4}}\right)\,.
\end{equation}

\section{Results and discussion \label{s-3}}

Here we report the results for the spectrum measured by observers at infinity for the two cases discussed above with a specific choice of the parameters involved. The first model is the homogeneous dust collapse (Oppenheimer-Snyder model) which terminates with the creation of a black hole. 
In this case the high density region that develops close to the formation of the singularity is always covered by the horizon. The second example is the inhomogeneous dust collapse model in which we find the formation of a singularity that can be visible to far away observers. The singular point from which null geodesics can propagate to infinity is that of $t_s(0)$ as given in equation \eqref{ts2}. All the other points in the singularity curve $t_s(r)$ are covered by the apparent horizon, nevertheless it can be shown that once there exist one radial null geodesic emanating from $t_s(0)$ that can reach observers at infinity then there exists a whole family of null geodesics and also non radial and non null geodesics can escape
(see for example Refs.~\cite{geodesics,geodesics2}).
Still, the amount of radiation that can escape from the high density region close to the singularity could in principle be very small and thus negligible in comparison with the spectrum emitted from the low density region which behaves similarly to the homogeneous cloud. This in fact turns out to be the case as the qualitative features of both spectrums result to be similar.

For the numerical calculations we have therefore set a scale by fixing the total mass of the object in both cases. More specifically, we have set the total mass $2M_{{\rm Sch}}=1$.
What changes is the way the mass is distributed within the cloud. Having fixed the value of the parameter $M_{0}=0.01$ to be the same in both models, in order to have the same total mass we must retrieve a different boundary in the two cases
\footnote{Another way of proceeding might have been to fix the same boundary for both models and thus have different values of $M_0$.}.
Then for the inhomogeneous collapse model we have set $M_{2}=-0.00015$. The parameter $M_2$ of the second model has been chosen, among all the allowed values for which the singularity is globally visible, in such a way to have the maximal effect in the difference between the two light curves. From the relation $2M_{{\rm Sch}}=r_b^{3}M(r_b)$, 
we find that the radius of the boundary $r_{b}$ is $\approx 4.64$ in the homogeneous model
and $\approx 6.06$ in the inhomogeneous one. Let us notice that, for the above
choice of $M_0$ and $M_2$, the maximum boundary radius for
the inhomogeneous collapse model is $\approx 6.32$
(the density at larger radii becomes negative and thus unphysical), 
and this also maximizes the difference 
in the light curves between the homogeneous and inhomogeneous collapse.  
The calculations have been performed from the distant observer (locate at the 
radius $R_* = 10^7$ in units $2M_{{\rm Sch}}=1$) backward in time to the
collapsing object. At any time as measured by the distant observer, the photon trajectory is 
characterized only by the impact parameter $b$. Geodesics in the exterior
Schwarzschild spacetime have been computed with a 4-th order Runge-Kutta 
method. As in the interior LTB spacetime we have a second order differential
equation, we decided to use the Runge-Kutta-Nystrom method~\cite{Lund2009} 
for the calculation of the trajectories.

\begin{figure}[hh]
\begin{center}
\includegraphics[width=7.8cm]{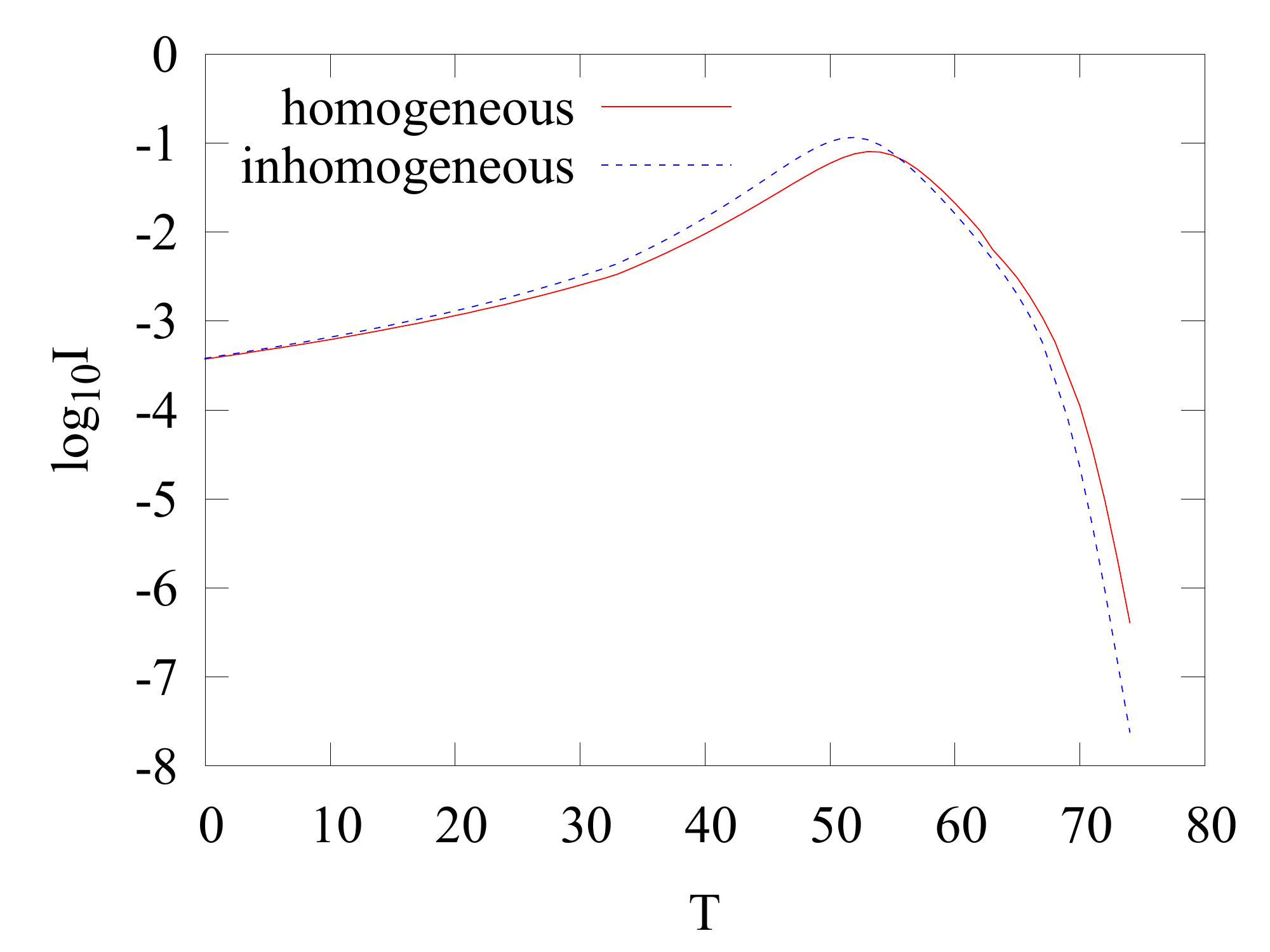}\hspace*{0.8cm}\includegraphics[width=7.8cm]{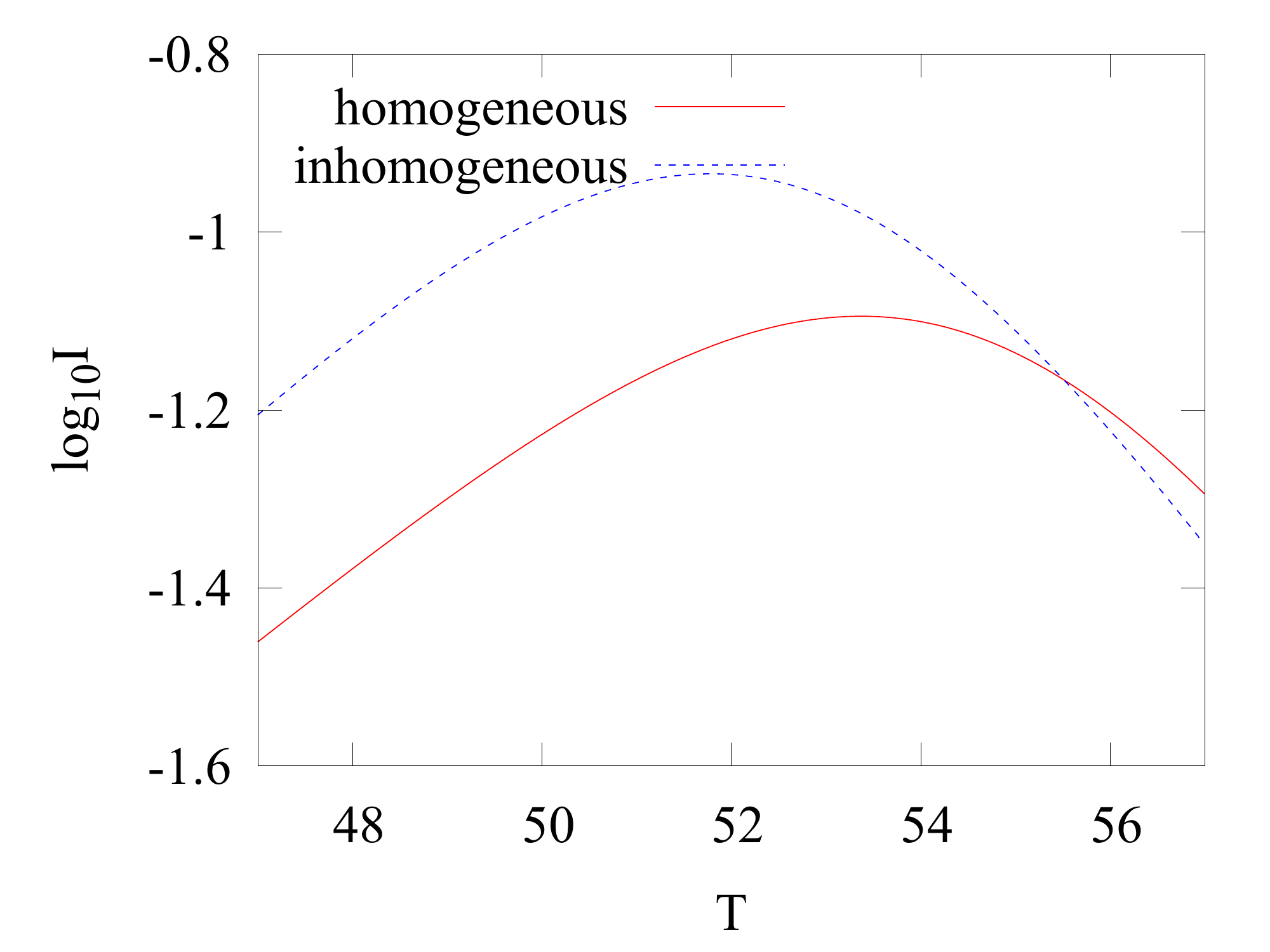}
\end{center}
\caption{Left panel: light curve luminosity of an LTB collapsing object with the 
emissivity function described by Eq.~(\ref{eq-delta}), for the homogeneous/black
hole (red solid curve) and the inhomogeneous/naked singularity case (blue dashed
curve). Right panel: zoom of the left panel at the peak of the luminosity.
$T$ in units $2M_{{\rm Sch}}=1$. Luminosity in arbitrary units. \label{f-delta}}
\end{figure}

\begin{figure}[hh]
\begin{center}
\includegraphics[width=7.8cm]{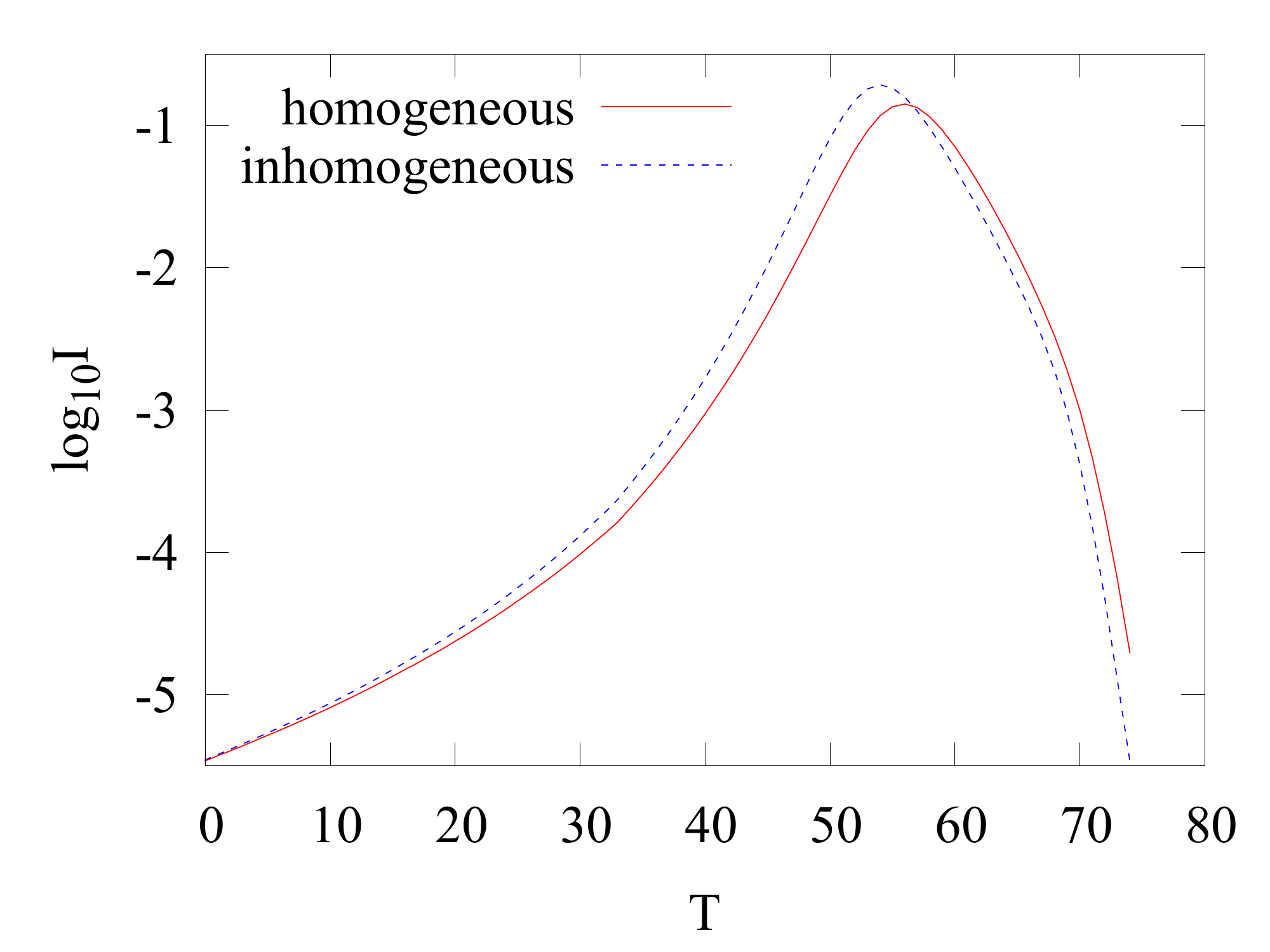}\hspace*{0.8cm}\includegraphics[width=7.8cm]{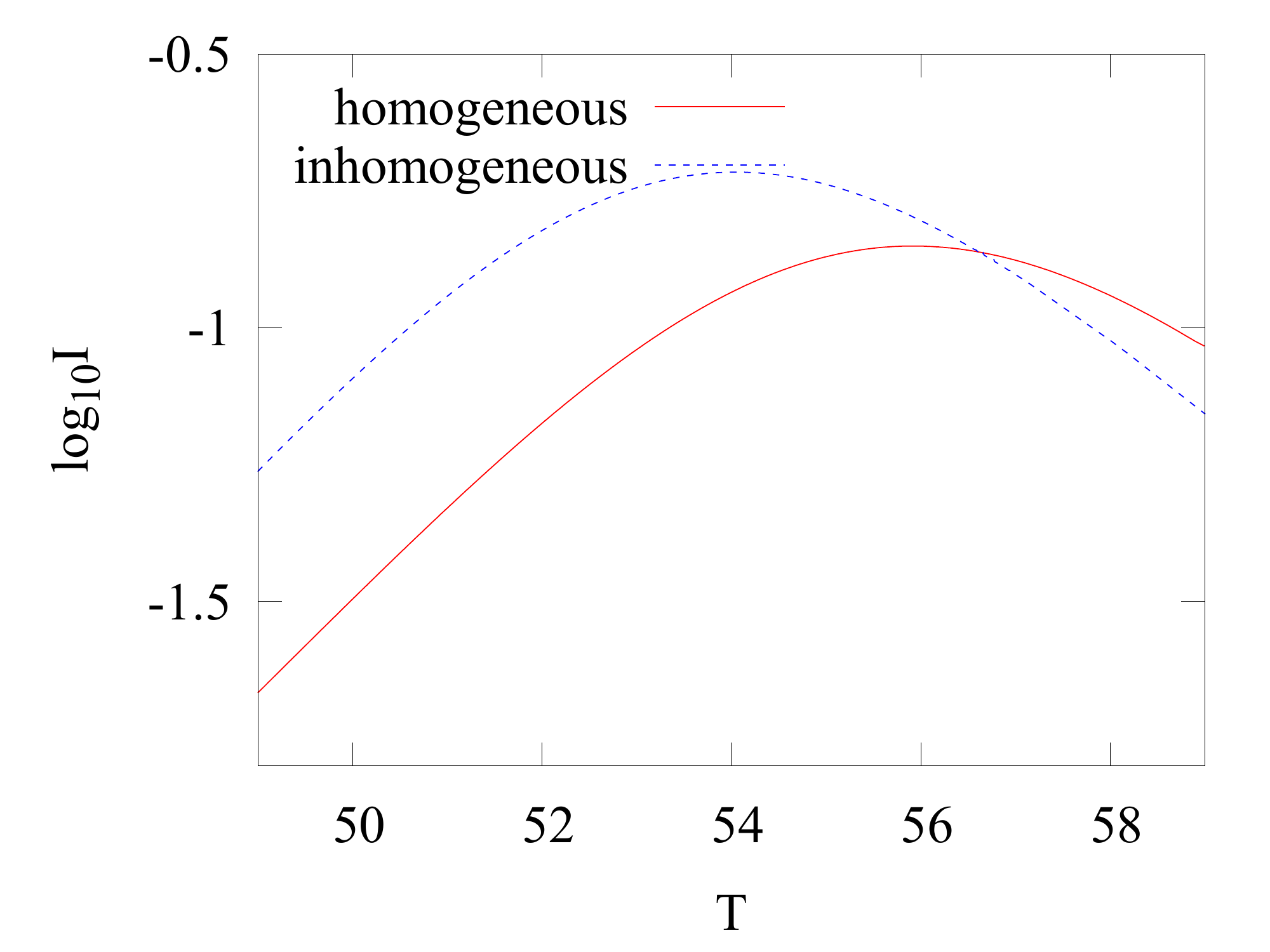}
\par\end{center}
\caption{As in Fig.~\ref{f-delta} for the emissivity function in Eq.~(\ref{eq-exp}).
$T$ in units $2M_{{\rm Sch}}=1$.
Luminosity in arbitrary units. \label{f-exp}}
\end{figure}

\begin{figure}
\begin{center}
\includegraphics[width=7.8cm]{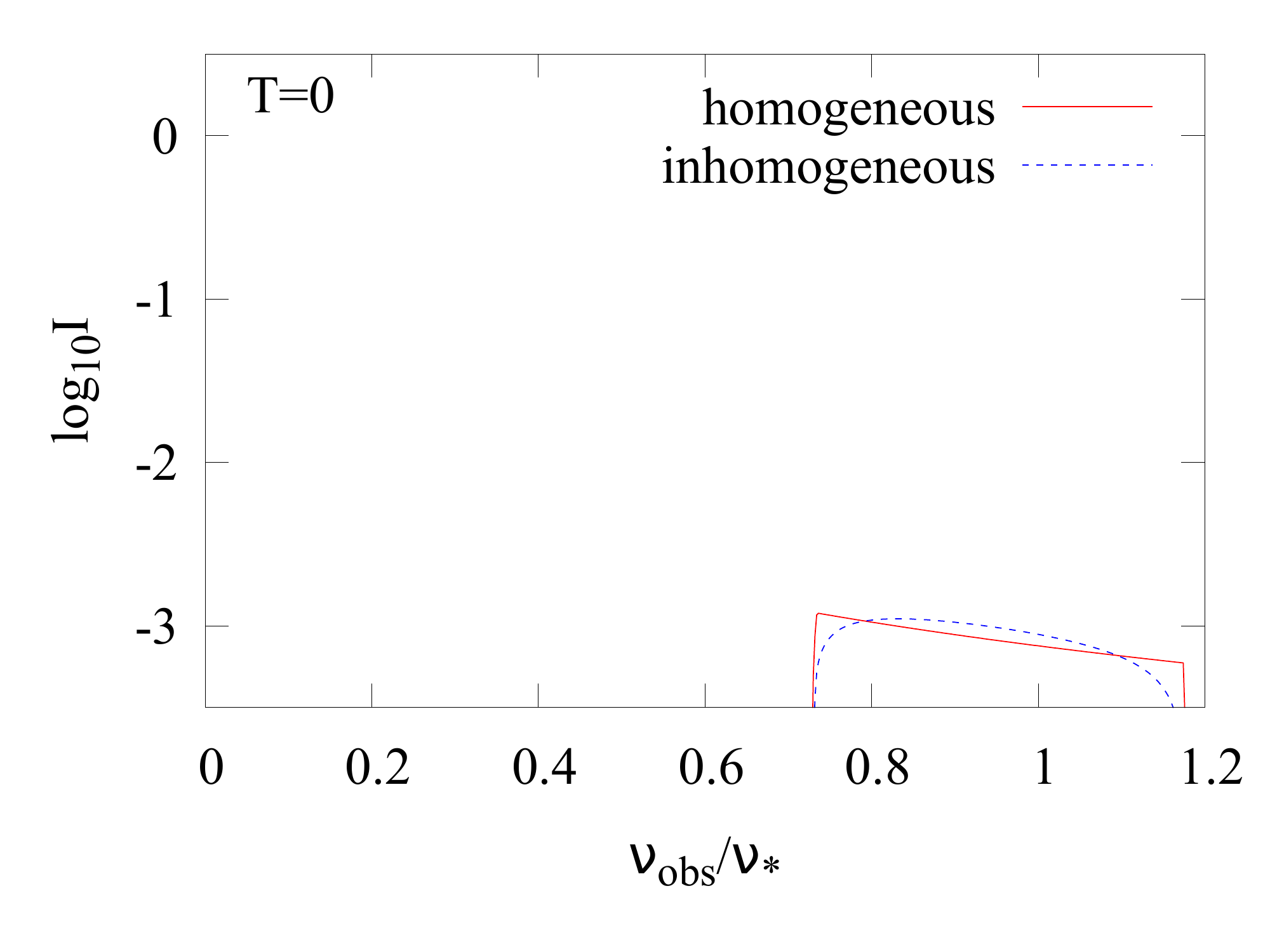}\hspace*{0.8cm}\includegraphics[width=7.8cm]{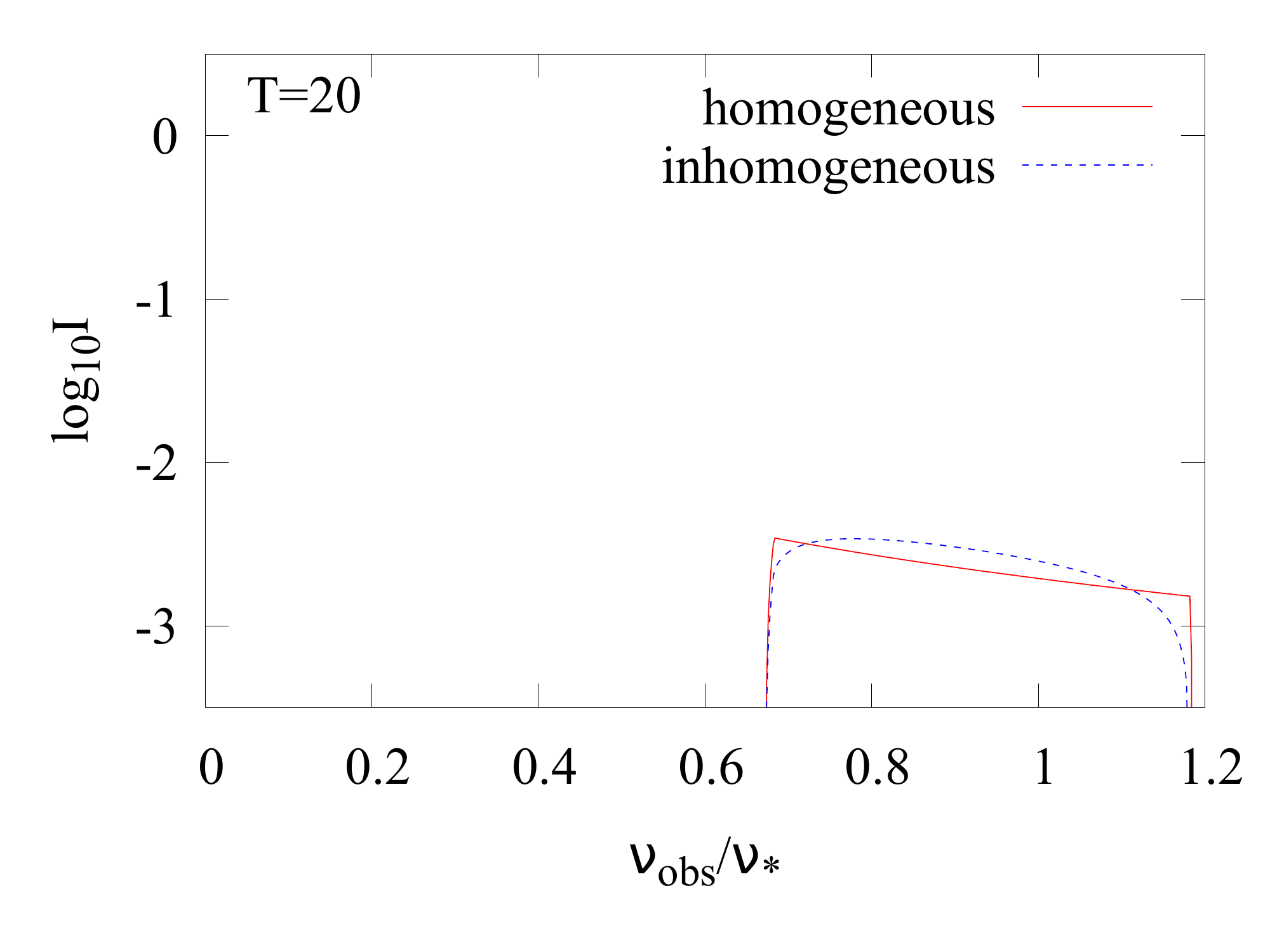}
\par\end{center}
\begin{center}
\includegraphics[width=7.8cm]{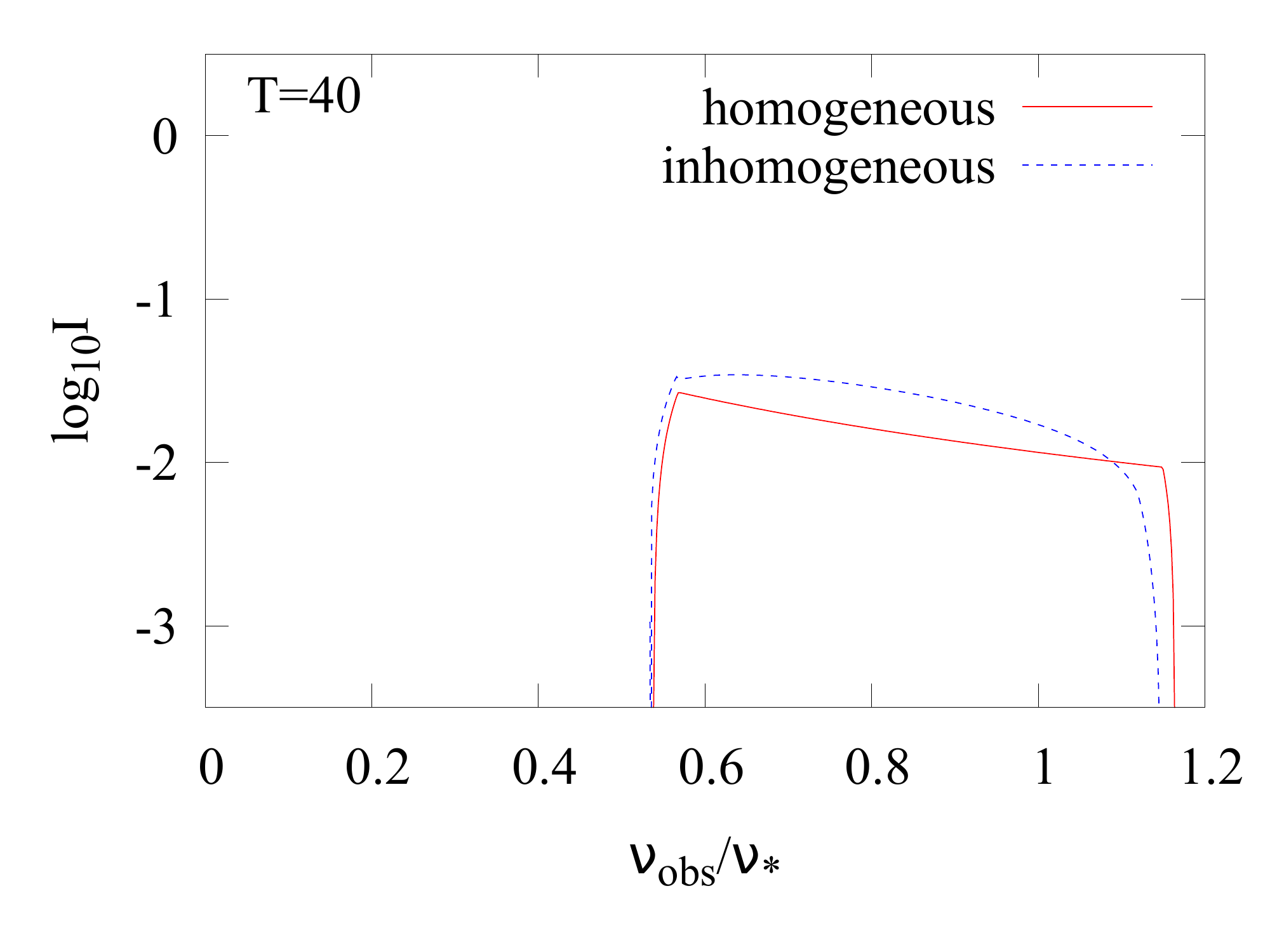}\hspace*{0.8cm}\includegraphics[width=7.8cm]{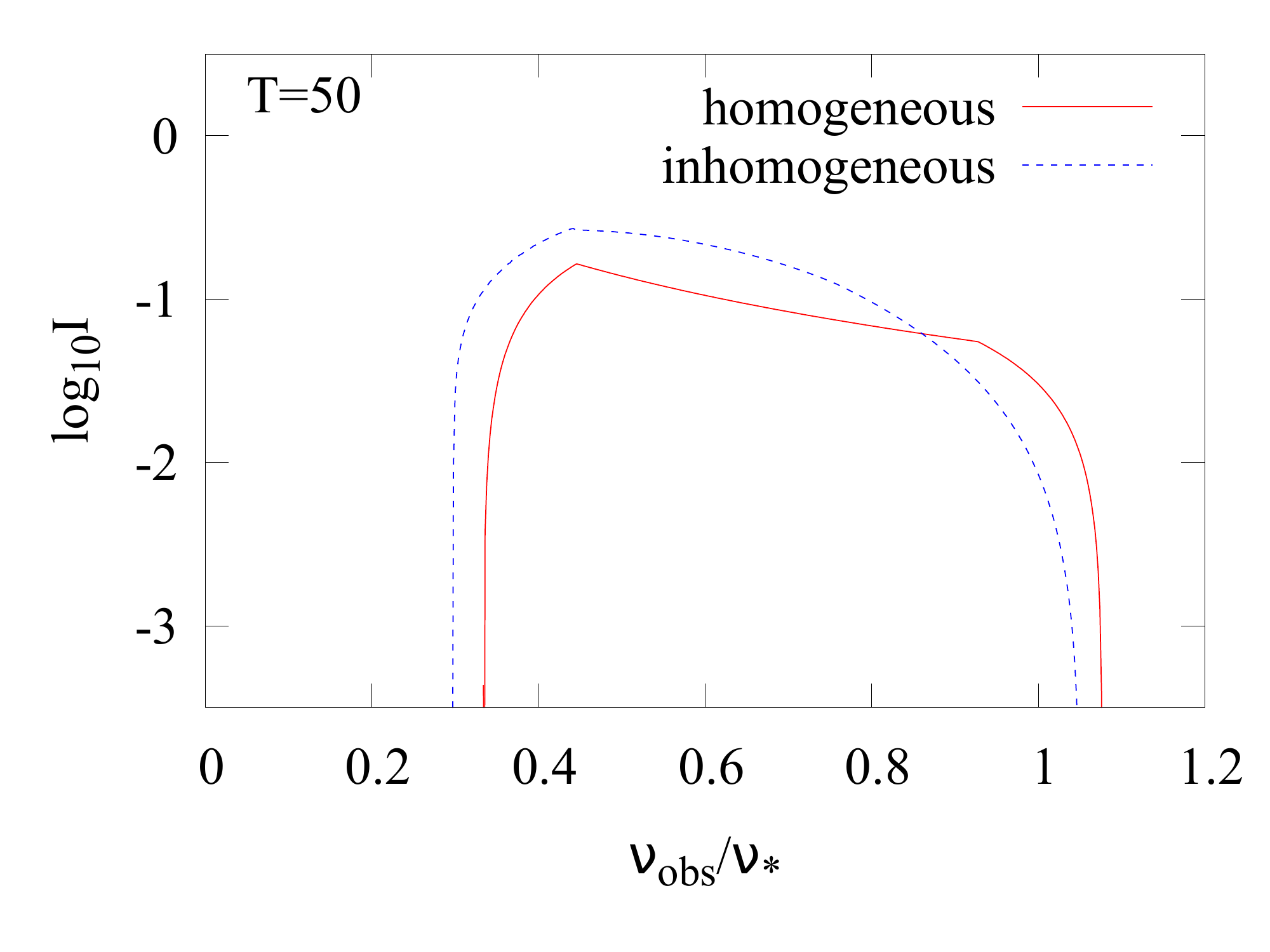}
\par\end{center}
\begin{center}
\includegraphics[width=7.8cm]{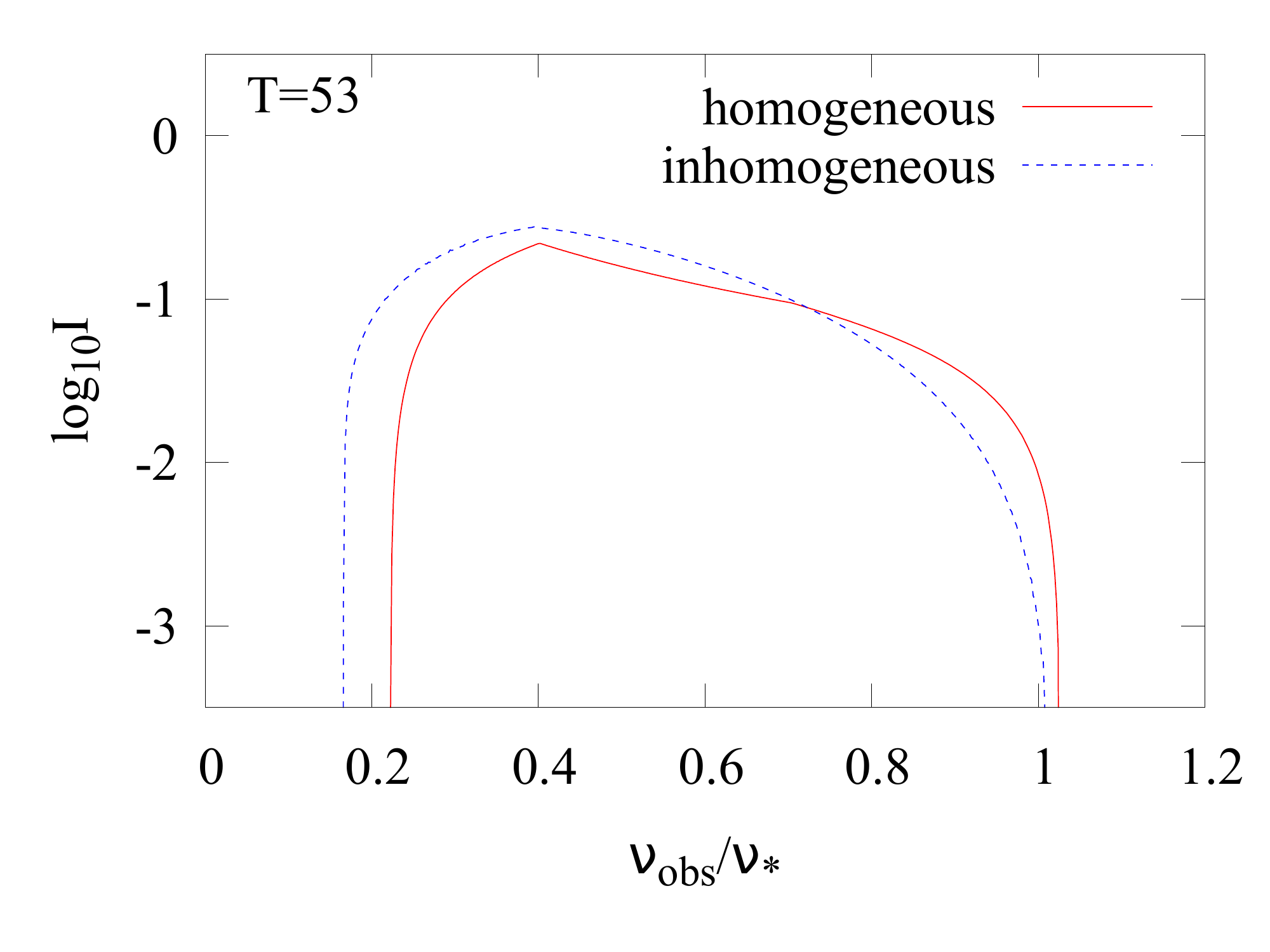}\hspace*{0.8cm}\includegraphics[width=7.8cm]{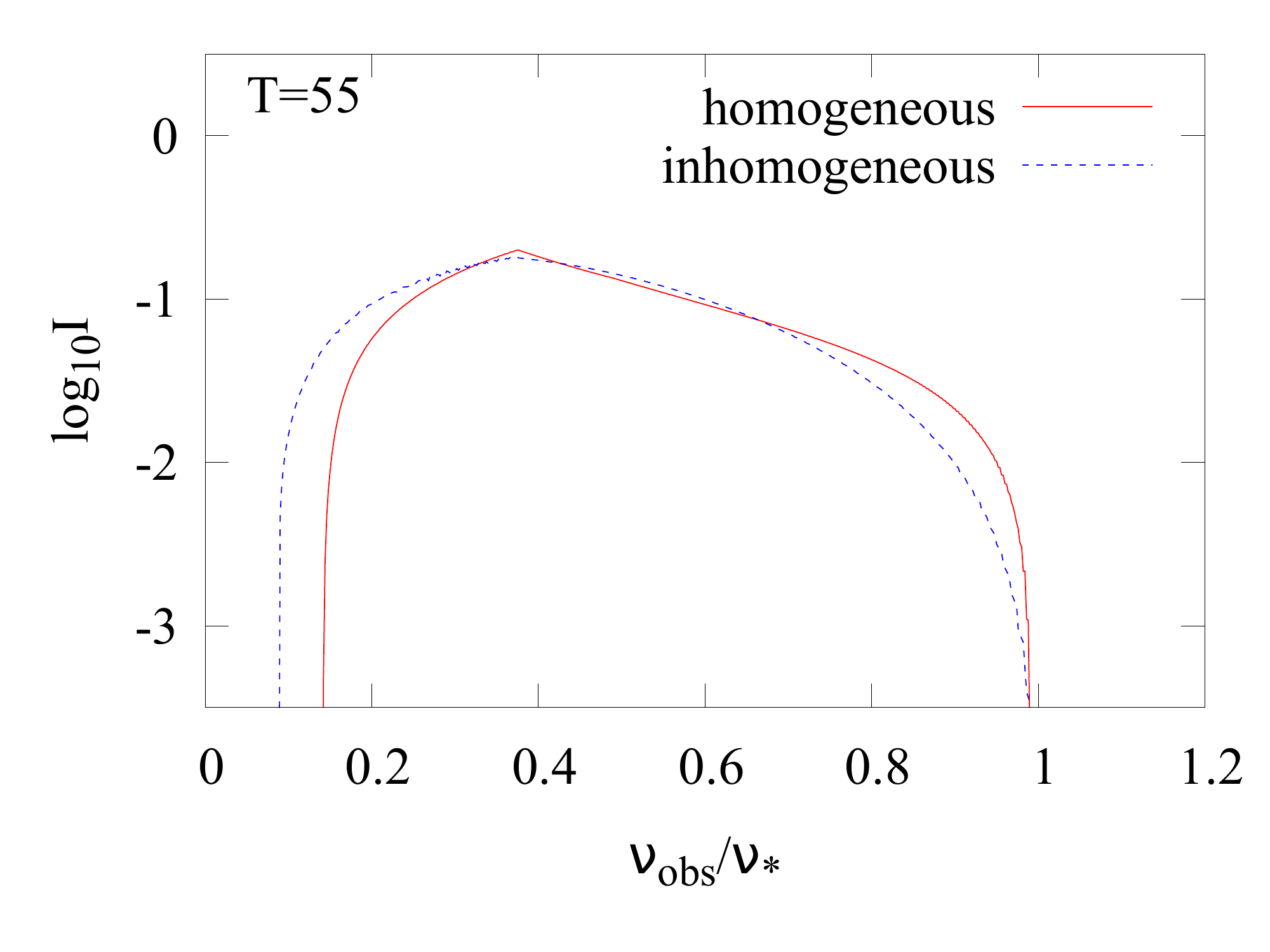}
\par\end{center}
\caption{Spectra of an LTB collapsing object with the emissivity function described 
by Eq.~(\ref{eq-delta}), for the homogeneous/black hole (red solid curve) and the 
inhomogeneous/naked singularity case (blue dashed curve).
Luminosity in arbitrary units. \label{f-s-delta}}
\end{figure}

\begin{figure}
\begin{center}
\includegraphics[width=7.8cm]{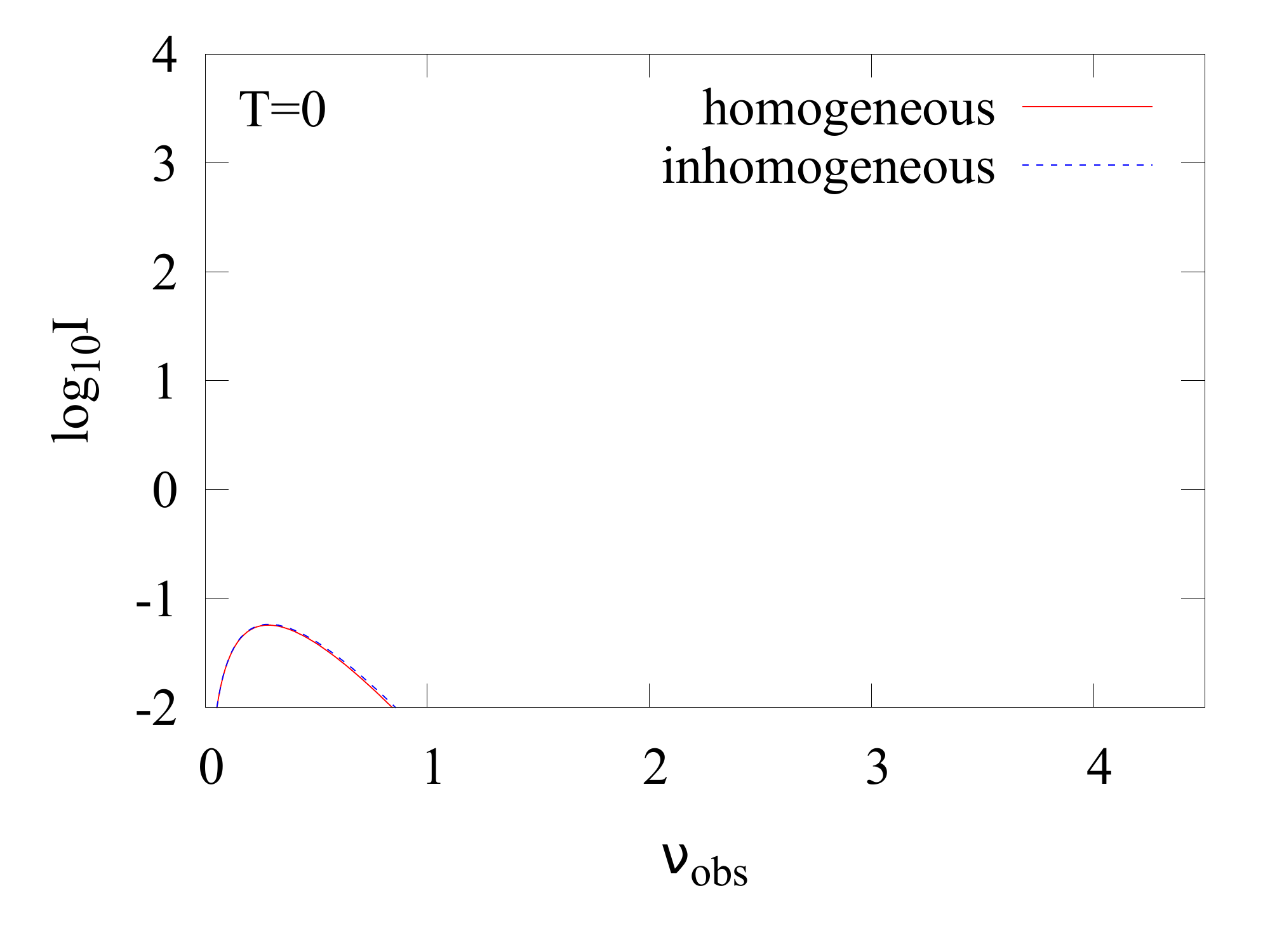}\hspace*{0.8cm}\includegraphics[width=7.8cm]{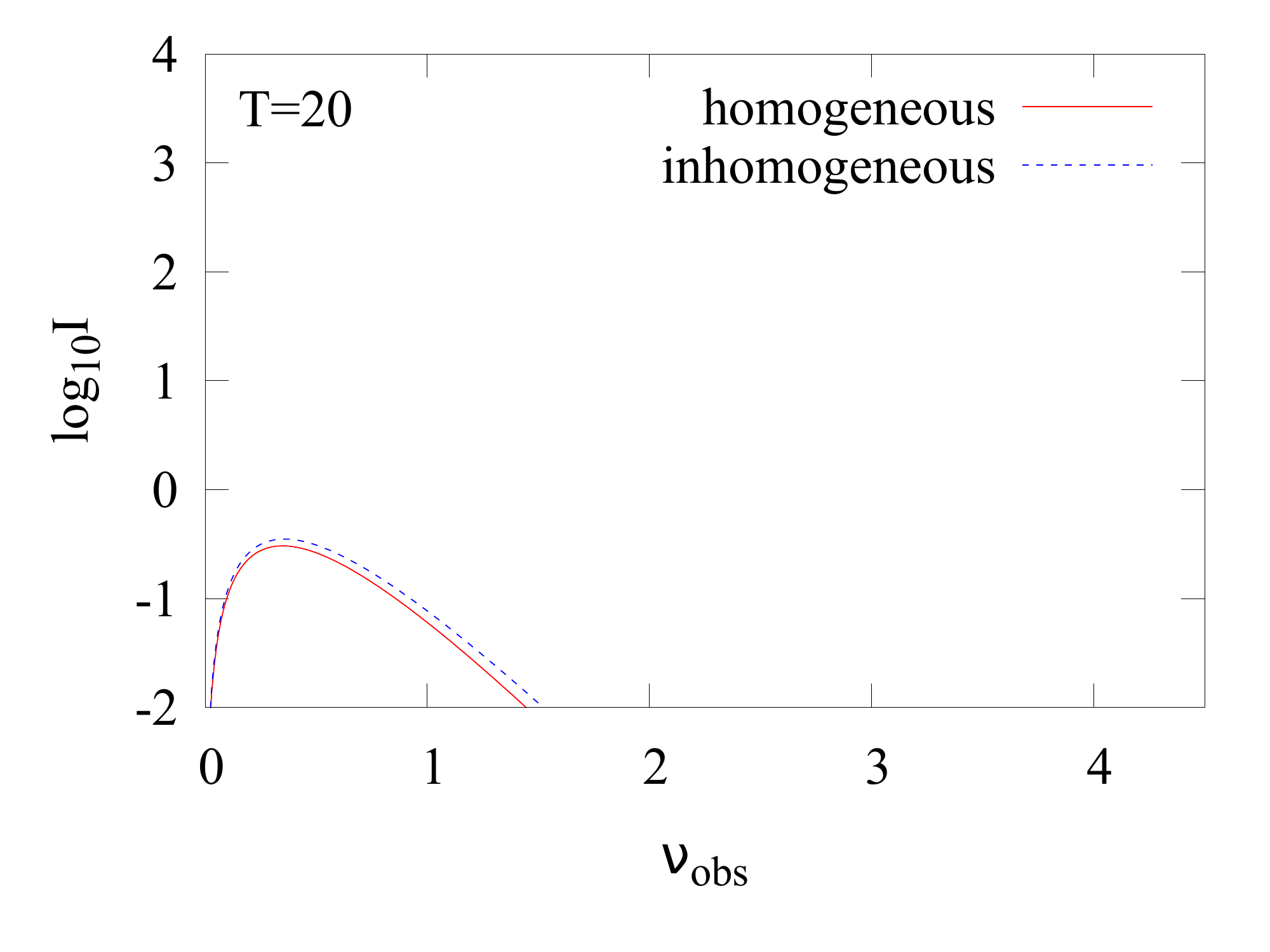}
\par\end{center}
\begin{center}
\includegraphics[width=7.8cm]{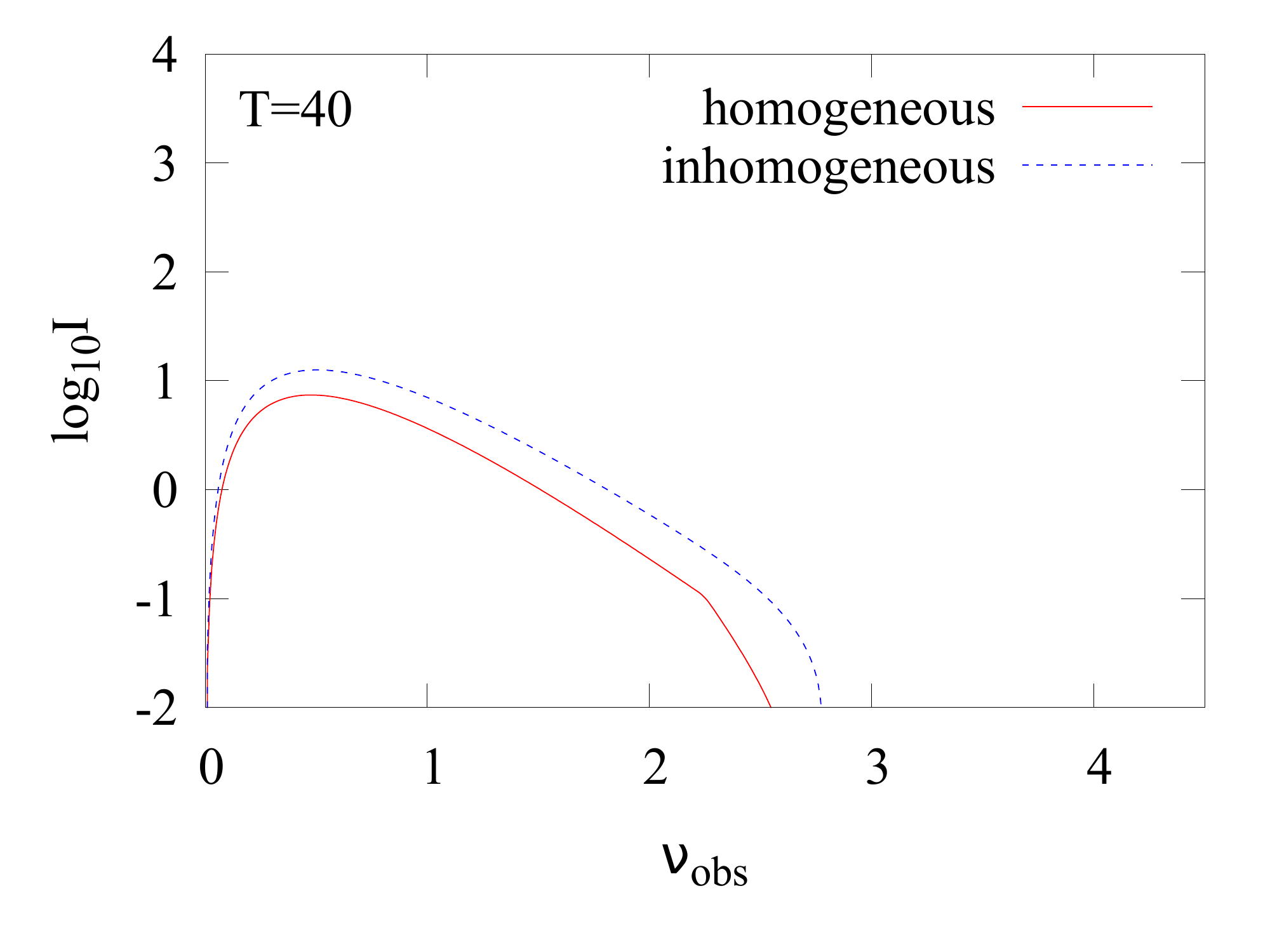}\hspace*{0.8cm}\includegraphics[width=7.8cm]{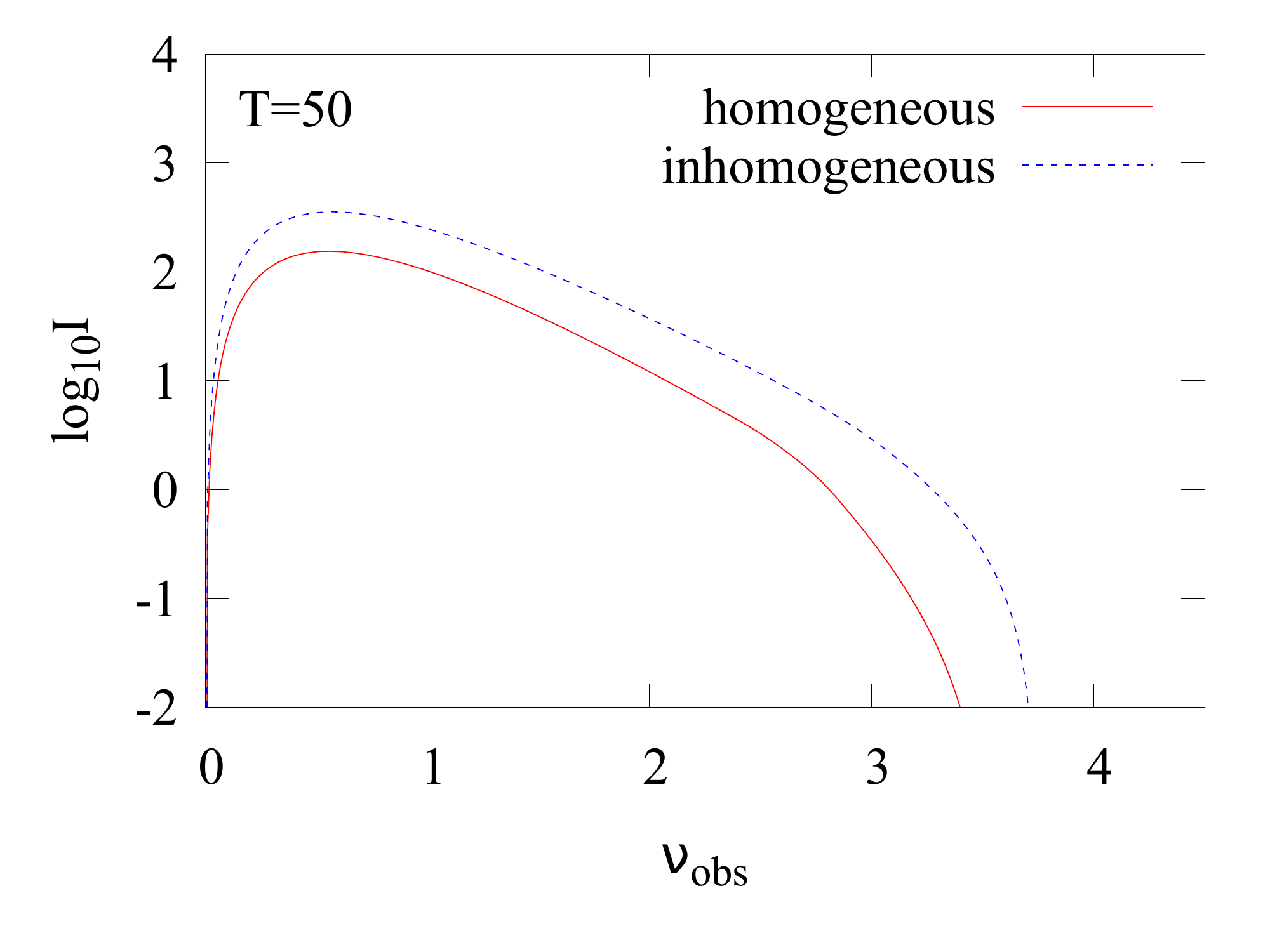}
\par\end{center}
\begin{center}
\includegraphics[width=7.8cm]{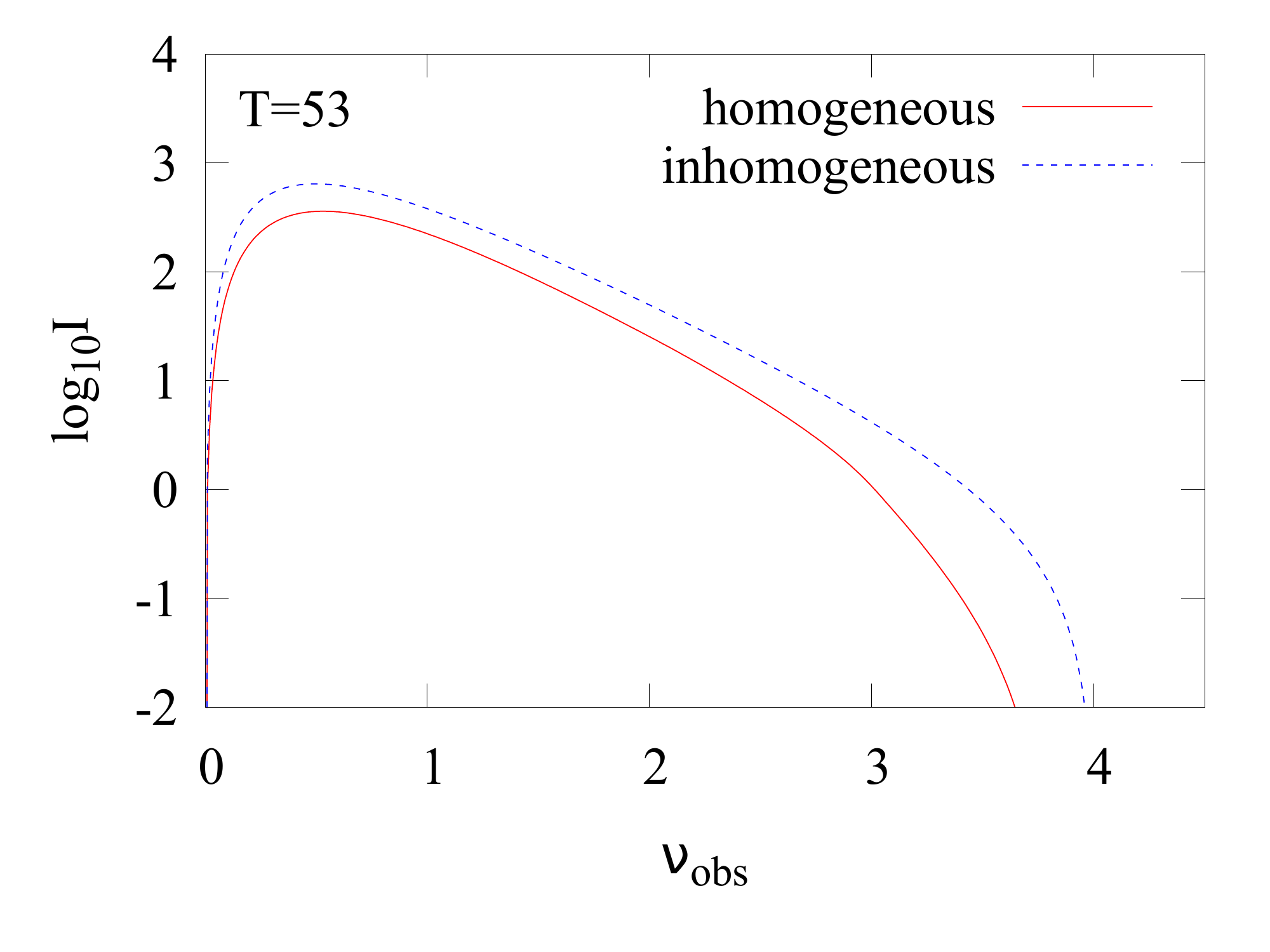}\hspace*{0.8cm}\includegraphics[width=7.8cm]{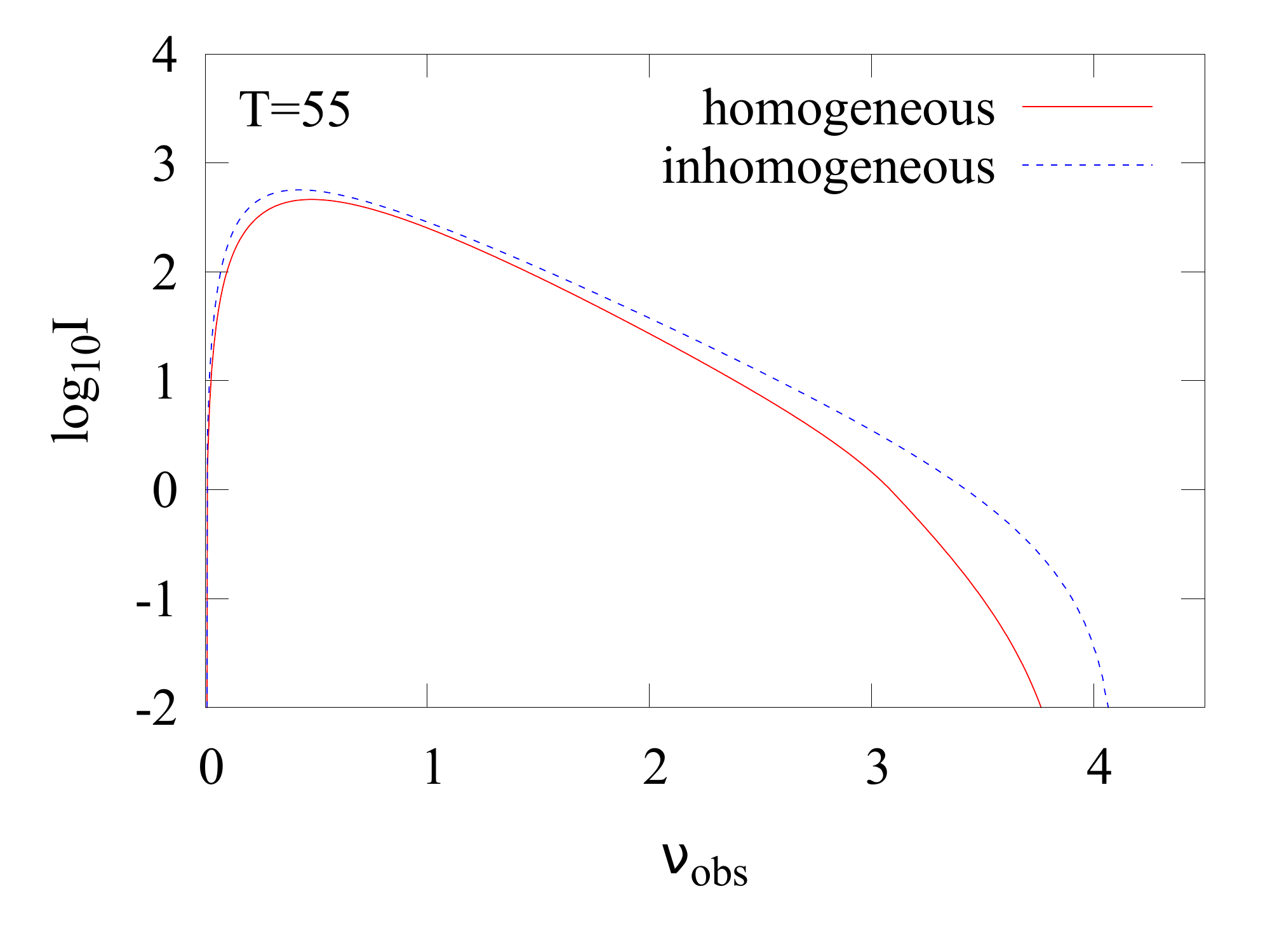}
\par\end{center}
\caption{As in Fig.~\ref{f-s-delta} for the emissivity function in Eq.~(\ref{eq-exp}). 
Luminosity and $\nu_{\rm obs}$ in arbitrary units. \label{f-s-exp}}
\end{figure}

The evolution of the light curve luminosity as a function of the proper time of
the distant observer $T$ for the homogeneous/black hole and 
inhomogeneous/naked 
singularity scenarios are reported in Figs.~\ref{f-delta} and \ref{f-exp},
respectively for the monochromatic emissivity of Eq.~(\ref{eq-delta}) and the thermal-like
one of Eq.~(\ref{eq-exp}). Figs.~\ref{f-s-delta} and \ref{f-s-exp} show instead some
spectra at specific times. The blueshift ($\nu_{\rm obs} > \nu_*$) shown in some
panels of Fig.~\ref{f-s-delta} is an effect experienced by those photons that can
propagate for a long time inside the collapsing object, so that the blueshift gained
in the interior solution exceed the redshift in the exterior Schwarzschild spacetime 
from the surface of the object to the distant observer. The phenomenon is
explained with some details in Ref.~\cite{kmb}.

As we can see from both the light curve luminosity and the spectra, there are no
qualitatively significant features that can differentiate between the two scenarios. 
While in the inhomogeneous/naked singularity case we could have expected a much
higher luminosity originating from the central region with high energy density
just before the formation of the naked singularity, it turns out the that size
of this region is too small, and the time scale too short, to produce 
a significant emission of radiation. On the one hand, this result may suggest
that the formation of a spacetime singularity visible to distant observers in 
our Universe is not a catastrophic event incompatible with observations.
It would seem that even within this simple and extremely idealized
collapse model it is impossible to observationally distinguish the birth of
a black hole from that of a naked singularity. 
Therefore, provided that the scenario is not drastically altered by some other effects 
(like for example those induced in the strong field regime by some theory of quantum gravity),
the creation of a spacetime naked singularity as the endstate of collapse might not have significant direct observational consequences for far away observers.
While we cannot at present exclude
the possibility of testing the weak cosmic censorship conjecture and/or
probe the high densities region where classically we would expect the formation 
of a spacetime singularity with astrophysical observations, it is clear that
such possibilities are at least challenging, even in principle, and are likely 
to remain challenging also in the case of more sophisticated theoretical models 
that account for more realistic scenarios and astrophysical effects.

\section{Summary and conclusions \label{s-c}}

In general relativity, gravitational collapse of type I matter fields satisfying basic energy conditions ends with the formation of a singularity of the spacetime, where the matter density diverges and standard physics breaks down. In particular this is the case for dust collapse, where, in the absence of pressures, a spacetime singularity is the only allowed outcome of collapse under the basic assumption of the positivity of mass and energy density.
Spacetime singularities may either be hidden behind a horizon, as in the case of black holes, or be naked and thus visible to distant observers. 
We now know many physically meaningful examples of naked singularities created as the endstate of collapse of matter fields that respect the standard energy conditions, starting with regular initial data. On the other hand, the weak cosmic censorship conjecture asserts that singularities produced in any generic gravitational collapse scenario must be hidden within black holes and cannot be seen by distant observers. The validity of this conjecture is still an open and controversial problem, but it is a key-assumption in black hole thermodynamics and it is of crucial importance for astrophysics where observed massive compact objects that exceed the Chandrasekhar mass limit are usually assumed to be black holes. 

In the present paper, we have tried to address the question whether it is possible to observationally test the weak cosmic censorship conjecture by measuring the radiation emitted by a collapsing body. 
In order to have a first understanding of the basic features of the problem, we decided to begin by studying the simplest theoretical collapse model, the LTB model, for which an analytical solution is known and easily calculated and to simplify as much as possible the assumptions related to the emitted radiation.
In this scenario the final product of collapse can be either a black hole or a naked singularity, depending on the values chosen for the parameters that determine the density profile. 
The naked singularity that forms as the endstate of dust collapse is naked only for a ``short time'' in comoving coordinates. Nevertheless this time may have been quite large once the photons emitted from the high density region reach observers at infinity.

We have computed the radiation emitted by these collapsing objects and their light curves, which can potentially track the evolution of the collapse, in order to find observational signature capable of distinguishing the birth of a black hole from the one of a naked singularity. Our collapse model is very simple and assume that the object is optically thin to the emitted radiation, which should make much easier the possibility of distinguishing the two scenarios than a realistic case with a lot of astrophysical complications. The answer to our question is not intuitively accessible, as the final result depends on several relativistic effects, like the gravitational redshift and the time delay between the collapsing star and the distant observer. Within our simple model, we did not find any specific signature to identify the naked singularity scenario. As shown in Figs.~\ref{f-delta} and \ref{f-exp}, the light curves for black holes and naked singularities do not seem to be qualitatively different. Roughly speaking, the reason is that the high density region formed just before the formation of the singularity is too small to produce an observational signature in the flux reaching the distant observer. While our finding cannot definitively exclude the possibility of observationally probing the high density region where classically we would expect the formation of a spacetime singularity, observational tests of the weak cosmic censorship conjecture seem to be at least extremely challenging, even in principle and even in the simplest case where we neglect all the possible astrophysical complications.


\begin{acknowledgments} 
This work was supported by the NSFC grant No.~11305038, the Innovation 
Program of Shanghai Municipal Education Commission grant No.~14ZZ001, 
the Thousand Young Talents Program, and Fudan University.
\end{acknowledgments}


\end{document}